\documentclass[%
 reprint,
superscriptaddress,
 amsmath,amssymb,
 pre,
]{revtex4-2}
\usepackage{centernot}
\usepackage{graphicx}
\usepackage{dcolumn}
\usepackage{bm}
\usepackage{times}
\usepackage{algpseudocodex}
\usepackage{algorithm}

\newcommand{\cut}[1]{{}}

\begin{document}


\title{Cellular Adaptation  to  Signal Fluctuations as  Learning}

\author{Tuan Minh Pham}
\email[Contact author: ]{m.t.pham@uva.nl}
\affiliation{Dutch Institute for Emergent Phenomena, 1090 GE, Amsterdam,
The Netherlands
}
\affiliation{Complexity Science Hub Vienna, Metternichgasse 8, A-1030, Vienna, Austria}
\affiliation{Institute for Advanced Study,
Oude Turfmarkt 147, 1012 GC Amsterdam, The Netherlands}
\affiliation{Institute of Physics, University of Amsterdam, Science Park 904, Amsterdam, The Netherlands}

\author{David Saad}
\affiliation{College of Engineering and Physical Sciences, Aston University, Birmingham B4 7ET, United Kingdom}
\begin{abstract}
 Cells represent one of the most fundamental units of life.  Underlying their robust performance against environmental variability, such as
temporal  fluctuations of  chemical signals, 
is a dynamical interrelation  between the two components of an intracellular pathway: a gene-regulatory network  and its upstream signal transducers. 
 To understand how   a \emph{single} cell utilizes this feedback to self-regulate its  gene-expressions, 
we develop a multiscale  model of the pathway's components, in which the \emph{adaptive} variables responsible for signal interpretation follow a  feedback-induced  learning process. We then derive a macroscopic theory capturing  the  covariations  between these components -- so-called  collective modes. 
Our theory shows how cells can achieve  robust output  against signal fluctuations via self-regulation rather than a simple  noise suppression. 
Such robustness  corresponds to
a 
transition from random- to   structured  collective modes  beyond   a critical adaptation  rate. 
\end{abstract}

\maketitle


\section{Introduction}  
In  development, cells make decisions based on the  guidance of  \emph{external} molecules known as morphogen. Morphogens regulate transcription either directly or through a specific
transducer, 
following a cascade of intracellular interactions -- a signaling pathway~\cite{Kicheva}. In this pathway, the interpretation of morphogen 
 is mediated by transcriptional effectors, whose states can vary substantially  in the presence of signal fluctuations (due to noise in production,  degradation and transportation of the morphogen~ \cite{Muller}). To maintain their 
 functionality, cells hence need to keep these effectors in a relatively stable state.

How cells can achieve this robustness through the course of development remains an open question~\cite{EBISUYA}.
At the molecular-level, cells integrate external signals to coordinate the action of multiple genes~\cite{Ham2025}.  Our understanding of such coordination has improved greatly thanks to recent advances in single-cell technologies~\cite{Tang2011}. In particular, cross-gene interactions in gene regulatory networks (GRNs) are now known in great detail. This opens up the possibility to study robustness against morphogen volatility by connecting molecular- and cellular-level descriptions of developmental processes~\cite{MacArthur, Teschendorff2021, GORIN2023822}. 

Following Waddington's idea~\cite{waddington1957}, GRN dynamics at the  cell level can  be  associated with a landscape  whose valleys correspond to 
different cell types~\cite{Rand2021} and which can be reconstructed  from single-cell data~\cite{Zhu,  Zhou2016, Zhou2021, Villarreal, Wang2011,Lang2014}. 
By deforming this landscape, morphogen guides the receiving cell to the appropriate attractors (cellular phenotypes), without allowing this cell to regulate the strength of the  signal~\cite{WOLPERT1969}. 

A commonly used example of this picture is the patterning of the vertebrate ventral neural tube, where the morphogen Sonic hedgehog (Shh) through its transcriptional effector Gli, specifies neural progenitor domains along the ventral-dorsal axis~\cite{Jessell}. Specifically, once  activated by the Shh gradient, Gli  controls the expression of downstream target genes (like Nkx2.2, Irx3, Olig2 and Pax6). These genes then regulate their own and each other's expression via a GRN. This network  settles into one of several discrete and stable attractors corresponding to distinct neuronal cell types. 

When the Shh concentration gradient constantly varies,  to maintain their type, cells need to actively regulate the interpretation of this variable signal by tuning Gli's activity~\cite{Cohen2015}. A previous study~\cite{Pezzotta} has
accounted for such adaptation mechanisms   by integrating   morphogen and  the GRN's landscape into   a new landscape. Despite its success, this approach ignores the non-equilibrium nature of GRNs due to \emph{asymmetric} interactions between genes as well as the \emph{timescale difference} between the signal  and that of the  GRN dynamics~\cite{BALASKAS} [the amount of Shh  in ventral cells changes over several hours, while the target genes  respond on a  faster timescale of tens of minutes]. It is well known that the potential-like landscape generally does not exist under these conditions~\cite{Guillemin}. Moreover,  while slow temporal signal integration has been reported as an important  mechanism for GRN adaptation~\cite{Dessaud2007}, timescale separation has so far been included to allow cells to track a slowly moving attractor, but is not considered a key factor allowing target genes  to determine their own attractors.

In this paper, we address these limitations by treating  the intracellular dynamics with two components operating on two different timescales. Specifically, in line with experiments~\cite{BALASKAS}, we assume that the target genes  follow stochastic dynamics that are much faster than those of the effectors. 
Formalizing  this latter  dynamics  as a non-equilibrium learning process \cite{engel2001statistical} of the adaptive variables describing the pathway’s effective sensitivity to signal fluctuations, 
we   capture the dynamical feedback between  these fast and slow processes at the microscopic level.  Our \emph{self-regulating} mechanism for robustness hence 
relies on the establishment of a correspondence between the two groups of transcription factors: the target genes and their  effectors.

To facilitate future comparison with   sequencing-based experimental studies~\cite{Gupta, Skinner, Olmeda2025}, where the collective modes of high-dimensional gene expression 
are typically quantified by  the covariations of different transcription factors,  we next  derive a closed set of dynamical equations for these covariances. Analyzing its numerical solutions,  we find that  (i)  robustness against morphogen fluctuations emerges beyond a critical    adaptation rate, and (ii) the attractor to be selected, which can be either fixed-point or oscillating ones, 
 depends on the rate at which the effect of externally varying signals is integrated into the fast dynamics of GRN through its  effectors.   



\section{The microscopic model}
\label{sec:model}

\begin{figure}
    \centering   \includegraphics[width=0.96\linewidth]{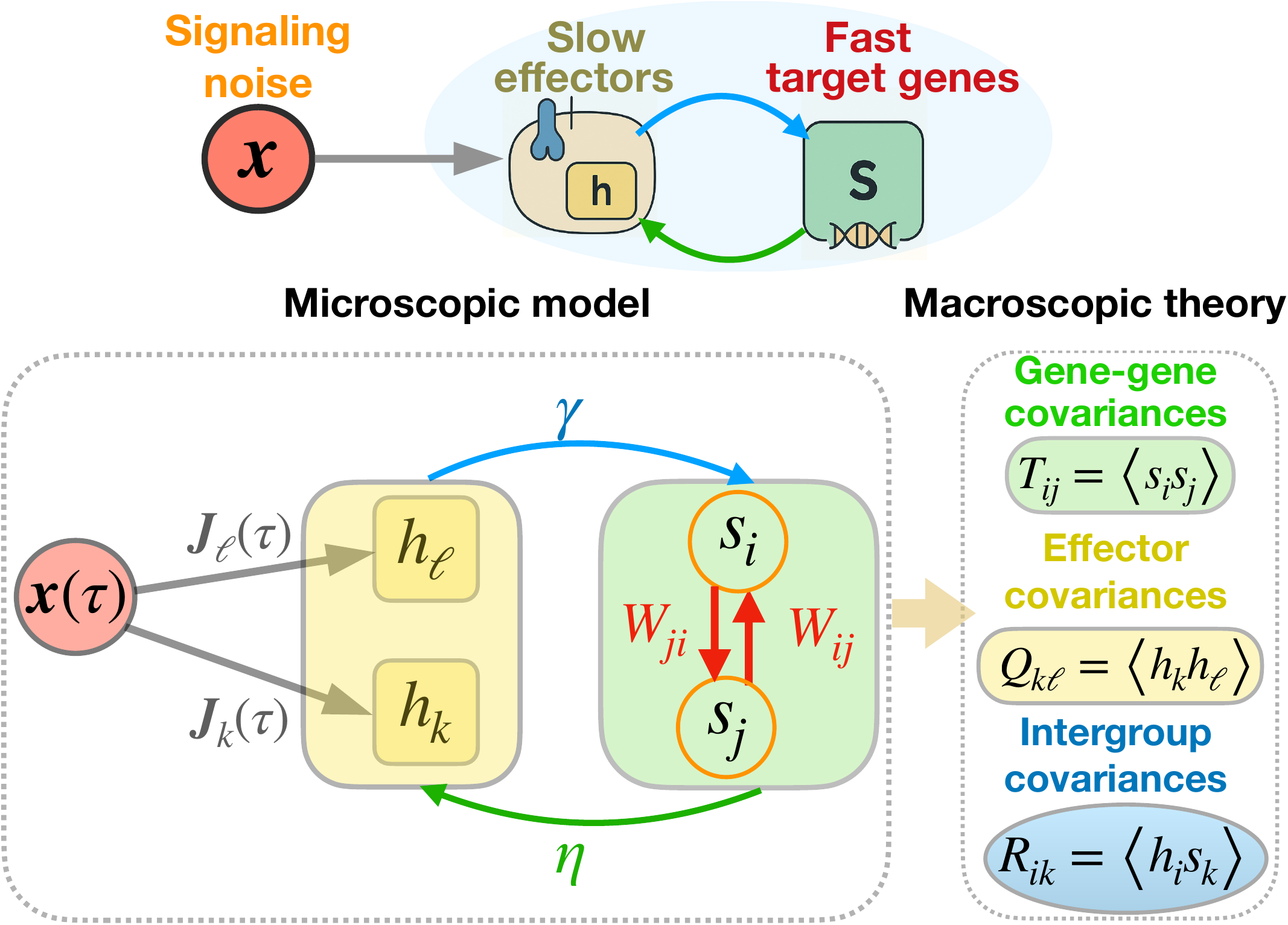}
\caption{\textbf{Top}. Illustration of an intracellular pathway  (marked by the shaded area) of effectors (labeled  by $\boldsymbol{h}$) and target genes (denoted by  $\boldsymbol{s}$). The pathway is subjected to  fluctuations of  
an external signal depicted by $\boldsymbol{x}$.  \textbf{Bottom}.
Our microscopic model~\eqref{learning_dynamics}-\eqref{fast_dynamics_tau}  and its corresponding  macroscopic theory in terms of collective covariances~\eqref{QRslow}-\eqref{Tslow}. Here  fluctuations    vary over a slow time $\tau$. The $M$ effectors   integrate these fluctuations into their states  $h_{\ell}(\tau)$ via an \emph{interpretation matrix} $J_{\ell \alpha}$ (denoted by black arrows):
$h_{\ell}(\tau) = \sum_{\alpha}J_{\ell \alpha} x_\alpha(\tau)$,
$\ell = 1, \cdots, M$ and $\alpha = 1, \cdots, N$. Blue arrow indicates  the effectors' controlling effect on the 
$K$ target  genes  with   strength $\gamma$, while green one with  $\eta$ depicts the upstream regulation of these genes on their effectors.  The target genes interact via a gene-regulatory network $\boldsymbol{W}$ with activating interaction (marked by red arrows). For illustration, we show the case of  $M=K =2$, but generally $M\neq K$.}
    \label{fig:schematic_pathway}
\end{figure}

Our model is based on the Shh pathway example discussed in the Introduction.   
This  pathway is characterized by signal transduction, transcription regulation and feedback as schematically depicted in Fig.~\ref{fig:schematic_pathway}. Here, the activities of upstream  effectors are directly controlled by the morphogen (input), while downstream target  transcriptional factors regulate each other as well as tuning the  effectors' state via  feedback. A cell does not directly control the signal, but can
maintain its state 
 by forming a closed-loop control between the  effectors and the target genes. Such self-regulation underlies the cell's  adaptive response to constantly variable input.

Let $\boldsymbol{x}^\mu= \big(x_1(\tau_\mu), x_2(\tau_\mu), \cdots, x_N(\tau_\mu)\big)^T$ 
denote the fluctuation of morphogen at discrete time points  $\tau_\mu$. Throughout this work, boldface symbols denote vectors and matrices, and the superscript $T$
denotes the transpose of a vector or a matrix. Here we consider $\tau_\mu = \mu/N$, where  $\mu=0,1,2\cdots \mu_{\rm max}$ with $\mu_{\rm max}$  the total number of time points and $N$ corresponds to the (large) number of signaling molecules.   Here, signal fluctuations refer to \emph{non-instructive}  stochastic temporal variations in signaling activity around a type-preserving input condition, rather than to instructive changes in morphogen exposure that are intended to drive a cell-type transition. Throughout this work, we consider $x_\alpha(\tau_\mu)$ as uncorrelated random variables with $\langle x_\alpha(\tau_\mu) \rangle = 0$ and $\langle x_\alpha(\tau_\mu) x_{\alpha'}(\tau_{\mu'}) \rangle = \delta_{\alpha \alpha'} \delta_{\mu \mu'}$. The cell  therefore does not  ignore genuine developmental information, but is able to actively regulate the interpretation of non-instructive fluctuations.

 The $M$ effectors  map (interpret) the state $\boldsymbol{x}^\mu$ onto their corresponding fields:
\begin{equation}
 h_k(\tau_\mu) := \sum_{\alpha=1}^N J^\mu_{k\alpha}x_\alpha(\tau_\mu) \,,\quad k=1,\cdots, M
 \label{h_definition}
\end{equation}
via a so-called \emph{interpretation matrix}  $\boldsymbol{J}^\mu$ consisting of  $J^\mu_{k\alpha}:=J_{k\alpha}(\tau_\mu)$, each maps the $\alpha$-th component of the fluctuating signal onto the $k$-th effector $h_k(\tau_\mu)$ at time $\tau_\mu$. Here,  $\boldsymbol{J}^\mu$ should be interpreted as a coarse-grained variable that represents 
the \emph{effective} sensitivity of the signaling pathway, 
but not as a set of  microscopic  couplings between the signal and the effectors. Nevertheless, for illustrating the model, in Fig.~\ref{fig:schematic_pathway} we depict   each $J^\mu_{k\alpha}$  as a  directed connection from $x_\alpha$ to $h_k$.  These effectors have  a response of  sigmoidal form
\begin{equation}
     g_k^\mu := \phi\big(h_k(\tau_\mu)\big)
 \label{phi_definition}
\end{equation}
where, throughout this paper, we shall use 
$$\phi(z)={\rm  erf}(z/\sqrt{2})= -1+(2/\pi)^{1/2} \int_{-\infty}^z du\exp(-u^2/2) ~.$$
 This activation function is chosen to  capture the threshold behavior of the Shh pathway~\cite{ Jessell}.  An effector $k$ is in its activator form ($g_k^\mu>0$) only
at sufficiently high level of $h_k(\tau_\mu)$, otherwise it is in the repressor form ($g_k^\mu<0$).


Next, we consider a  gene regulatory network $\mathcal{G}$ of $K$  target genes $s_i$, $i=1,\cdots, K$. For simplicity, we choose $M=K$, focusing on  effectors that directly affect target genes. 
The case of $M\neq K$ can be easily accommodated by adding additional weights \cite{Goldt2020}. The off-diagonal elements $\{W_{ij} \}_{i,j=1,\dots,K}$ of the weighted interaction matrix $\boldsymbol{W}$  associated with $\mathcal{G}$ encode the regulatory  interactions  between
pairs $(i,j)$ (if $i$ and $j$ do not
interact, then $W_{ij}=W_{ji}=0$). In general, $W_{ij}\neq W_{ji}$. As we do not consider self-interaction, the diagonal elements of $\boldsymbol{W}$ are zero.  Under fixed $g^\mu_i$, the dynamics of $s^\mu_i(t):= s_i(t;\tau_\mu)$ progress on a fast timescale $t$ compared to that of the effectors and are subjected to zero-mean Gaussian white noises $\xi^\mu_i(t):= \xi_i(t;\tau_\mu)$, $\langle \xi^\mu_i(t) \xi^{\nu}_{j}(t')\rangle = \sigma^2\delta_{\mu\nu} \delta_{ij} \delta(t-t')$ \cite{YUAN2021, Paczko, REINITZ1995, Matsushita2020}:
\begin{equation}
\frac{ \partial}{\partial t}  s^\mu_i(t) =  -s^\mu_i(t) +  c \phi\left(\sum_{j=1,j\neq i }^K  W_{ij}s^\mu_j(t)\right) + \gamma g^\mu_{i} +   \xi^\mu_i(t) \label{fast_dynamics}
\end{equation}
 Note that this equation has been nondimensionalized so that the decay term is given by $-s_i$. The parameter  $c$ hence  measures how strong the cross-regulations between target genes  is relative to their intrinsic relaxation, while  a coefficient $\gamma$ encodes the cis-regulatory/transcriptional efficacy. Through $\gamma$ we add a control term  linear in $g^\mu_i$, which are  correlated among multiple genes. This term can
 enhance or suppress the expression of gene $i$ depending on the state of the effector, playing the role  of an attractor selection through the effectors. 
The additivity of the control is somewhat artificial and would generally not apply if we were to derive our model from a microscopic model of cis-regulations \cite{Sherman, Goutsias}. However, this allows us to separate the dynamics into intra-system control within $\mathcal{G}$ and external driving by the  effectors, illustrating better the role of self-regulation. Here,  for simplicity, we consider additive noise, although multiplicative noises such as in ~\cite{COOMER202283} can be treated as well. The fast dynamics in Eq.~\eqref{fast_dynamics} typically have  multiple  attractors ~\cite{Sagner2017}~\footnote{In biological terms, such  multistability corresponds to the induction
of a distinct set of target genes in different tissues in
response to the same signal -- a context-dependent interpretation}. A sufficiently large  noise $\boldsymbol{\xi}$  can then result in stochastic switchings between alternative attractors. Since we are not interested in how gene-expression noise affects   switching probabilities in this paper, we set $\sigma=0.01$. 

Due to variations of the morphogen level $\boldsymbol{x}^\mu$, under fixed  $\boldsymbol{J}^\mu = \boldsymbol{J}^{\mu +1}$, $h_k(\tau_\mu)$ and hence $g_k^\mu$ 
may change significantly between time steps, from $\tau_\mu$ to $\tau_{\mu+1}$. As a result, the cell cannot maintain a robust gene-expression pattern of its target genes if $\boldsymbol{J}$ is \emph{quenched} (in its contribution to Eq.~\eqref{fast_dynamics}). To understand  the cell's ability to maintain   a certain phenotype, 
despite morphogen fluctuations, we propose to consider $\boldsymbol{J}$ as dynamical variables that evolve over time $\tau_\mu$. Moreover, the update of $\boldsymbol{J}^\mu$ to $\boldsymbol{J}^{\mu+1}$ aims at reducing 
the following mismatch between the target genes' total output $Z^\mu$ and that of all effectors $Y^\mu$ [with the latter determined by $\boldsymbol{x}^\mu$]:
\begin{equation}
\epsilon(\tau_{\mu}):=\epsilon(\boldsymbol{x}^\mu, \boldsymbol{J}^\mu) = \frac{1}{2}\, \big(Y^\mu - Z^\mu\big)^2 
\label{generation_error_in_dynamic_networks}
\end{equation}
where 
\begin{equation}
    Y^\mu:= \sum_{\ell=1}^K \phi(h_\ell(\tau_\mu))\,,\qquad  Z^\mu:=\sum_{i=1}^K   \phi\big(s_i^\mu(t\rightarrow \infty)\big)
    \label{output_definition}
\end{equation}
Therefore, the minimization of $\epsilon$ can be  understood
as a fluctuation-suppressing mechanism which arises due to correlation between
 the target genes and their effectors. 
 One would expect  an asymptotically diminishing value  $\epsilon(\tau_{\mu}\rightarrow \infty)\rightarrow 0$ as the cell commits to a give type at steady state. 
 To this end, we introduce  a so-called adaptation rate $\eta$ and  describe the adaptive response of the signaling pathway by the following  gradient-descent dynamics for $J_{k\alpha}(\tau_\mu)$:
\begin{equation}
J_{k\alpha}(\tau_{\mu+1})=J_{k\alpha}(\tau_{\mu})- \frac{\eta}{N}\frac{\partial \epsilon(\tau_\mu)}{\partial J_{k\alpha}(\tau_\mu)} \label{discrete_learning_dynamics}
\end{equation}
Hence $\eta$ modulates the sensitivity of the pathway. 
When $\eta=0$, the feedback between the two dynamics is turned off. In experiment,  $\eta \propto 1/\tau_0$, where $\tau_0$ is  the  timescale for  temporal integration of morphogen~\cite{Dessaud2007}.

The interlinked dynamics in Eqs.~\eqref{fast_dynamics}-\eqref{discrete_learning_dynamics} constitute our system of interest. We  implement this  set of equations in a nested fashion. Specifically, at time $\tau_\mu$, we integrate Eq.~\eqref{fast_dynamics}  under the present value of $\boldsymbol{g}^\mu$ until  a steady-state solution  $\boldsymbol{s}^\mu(t\rightarrow\infty)$ is obtained. This solution then is used to compute $Z^\mu$ and $\epsilon(\tau_\mu)$. Consequently  we update $\boldsymbol{J}^\mu$ to $\boldsymbol{J}^{\mu+1}$ according to Eq.~\eqref{discrete_learning_dynamics}. This procedure is again repeated at $\tau_{\mu+1}$. Underlying this implementation is our  assumption that the duration between $\tau_{\mu}$ and $\tau_{\mu+1}$ must be long enough  for the  fast dynamics to relax to its asymptotic attractor, which is a steady state defined by the controller $\boldsymbol{g}^{\mu}$  at $\tau_{\mu}$.

The iterative steps of finding successive attractors correspond to the experimental observation~\cite{Dessaud2007} where  instead of committing to an attractor based on a single morphogen levels ``snapshot", a cell integrates the signal over time, ensuring that only sustained inputs  can change its identity. In this regard, the temporal integration of Shh~\cite{Dessaud2007} is what sets the timescale separation in our model.

In essence, our scheme relies on the \emph{structural stability} of the GRN  dynamics, according to which a  small change of $\boldsymbol{g}^\mu$ from  $\tau_\mu$ to $\tau_{\mu+1}$ still gives rise to a stable fixed point $\boldsymbol{s}^{\mu+1}(t\rightarrow\infty)$. In fact, such a change is small as  $g_k=\phi(h_k)$ 
and the change of $J_{k\alpha}(\tau_\mu)$ is of order $O(N^{-1})$, following Eq.~\eqref{discrete_learning_dynamics},  but has a cumulative effect on the effectors. This structural stability is, however,
not guaranteed after many iterations of the slow dynamics, preventing 
the
 GRN dynamics from reaching any stable equilibria at $\mu_{\rm max}\gg 1$. Therefore, we will restrict our study to  the region of the parameter space, where the fast dynamics is assumed to always converge to a stable fixed point (but not limit cycles or chaotic attractors, which could happen in principle under strong non-linearity and the asymmetry of $\boldsymbol{W}$). As we will
show in section~\ref{sec:results}, our results confirm the existence of such a region, where slow macroscopic   variables indeed  relax to stable patterns. 

\section{The macroscopic theory}
\label{sec:macrotheory}

 In the limit of an infinite number of signaling molecules $N\rightarrow \infty$, $\tau_\mu= \mu/N$ becomes a continuous-time variable $\tau$.
Taking this limit allows us to cast Eq.~\eqref{discrete_learning_dynamics} into an ODE form
\begin{equation}
\begin{aligned}
    \frac{d}{d\tau} \, J_{k\alpha}(\tau) &= \eta\, w_{k}(\tau) x_\alpha(\tau)
  \\ w_k(\tau) & := \phi'\big(h_k(\tau)\big) \big(Z(\tau)- Y(\tau)\big)
\label{learning_dynamics}
 \end{aligned}
\end{equation}
with  $\boldsymbol{h}(\tau), \boldsymbol{x}(\tau), \boldsymbol{J}(\tau):= \lim_{N \rightarrow \infty} \boldsymbol{h}(\tau_\mu),\boldsymbol{x}(\tau_\mu),\boldsymbol{J}(\tau_{\mu})$  and $\langle x_\alpha(\tau) x_{\alpha'}(\tau') \rangle = \delta_{\alpha \alpha'} \delta(\tau -\tau')$. This equation indicates that changes in the effective sensitivity are proportional to the product of mismatch $Z-Y$ , effector responsiveness $\phi'(h_k)$ and signal fluctuations $x_\alpha$. For the sake of clarity,  let us rewrite the fast dynamics in Eq.~\eqref{fast_dynamics} in terms of $t$ and $\tau$, and $\boldsymbol{s}(t;\tau):=\lim_{N \rightarrow \infty} \boldsymbol{s}^\mu(t)$: 
 \begin{equation}
\frac{ \partial}{\partial t}  s_i(t;\tau) =  -s_i(t;\tau) +  c \phi\left(\sum_{j=1}^K  W_{ij}s_j(t;\tau)\right) + \gamma g_{i}(\tau) +   \xi_i \label{fast_dynamics_tau}
\end{equation}
The  microscopic model comprising   Eqs.~\eqref{learning_dynamics}-\eqref{fast_dynamics_tau}  are difficult to analyze. In the limit $N\rightarrow \infty$, we can employ  self-averaging macroscopic variables instead. 
Firstly, we assume that the GRN settles into a quasi-steady state for each slow value of $\boldsymbol{g}$ as  on the fast timescale of $\boldsymbol{s}$, $\boldsymbol{g}$ is just a constant drive. We denote such \emph{fluctuating} state of target genes  (due to noise $\boldsymbol{\xi}$):
\begin{equation}
 s_i(\tau):=\lim_{t\rightarrow \infty}s_i(t;\tau)    ~.
 \end{equation}
 With a \emph{linear} approximation of their dynamics, that holds true at steady state,  the expression profile $\boldsymbol{s}(\tau)$ is approximately  a multivariate Gaussian~\cite{Saez2022}. We hence can characterize their states by the correlations:
 \begin{equation}
\boldsymbol{T}_\tau\big(t):=  \langle \boldsymbol{s}(t;\tau)\boldsymbol{s}^T(t;\tau)\rangle- \big\langle \boldsymbol{s}(t;\tau)\big\rangle\big\langle \boldsymbol{s}^T (t;\tau)\big\rangle    
 \end{equation}
 Let $\boldsymbol{A}:=-\mathbb{I}+ c\boldsymbol{W}$, where $\mathbb{I}$ is the identity matrix. Following~\cite{Godreche_2019}, for $\boldsymbol{A}$ being a Hurwitz stable matrix (i.e. all of its eigenvalues have negative real parts), we  obtain the equation of motion for the target genes'  covariance matrix 
 $\boldsymbol{T}_\tau(t)$:
 \begin{equation}
     \frac{d}{dt}\,  \boldsymbol{T}_\tau = \sigma^2\mathbb{I} + \boldsymbol{A}\boldsymbol{T}_\tau + \boldsymbol{T}_\tau\boldsymbol{A}^T +\gamma\big\langle\boldsymbol{g}(\tau)\boldsymbol{s}^T(t;\tau) + \boldsymbol{s}(t;\tau)\boldsymbol{g}^T(\tau)
      \big\rangle 
     \label{Tslow1}
 \end{equation}
The steady state of this dynamics,    $\boldsymbol{T}_\tau^{(\infty)}:=\lim_{t\rightarrow \infty} \boldsymbol{T}_\tau(t) $  can be considered as a  cell's phenotype because it corresponds to fluctuations
around a stable attractor of the fast dynamics represented by Eq.~\eqref{fast_dynamics_tau}.  For notational simplification, from now on we use $\boldsymbol{T}(\tau)$ instead of $\boldsymbol{T}_\tau^{(\infty)}$.
 
 Since $x_\alpha(\tau)$ are uncorrelated random variables, the effector fields $h_k(\tau)$  are  Gaussian  of zero mean and covariance
(this will not  hold true in general for arbitrary $\boldsymbol{J}(\tau)$ as remarked in~\cite{Mace1998}), such that:
\begin{equation}
    Q_{k\ell}(\tau) := \big\langle h_k(\tau) h_\ell(\tau)\big\rangle_{\boldsymbol{h}}=\sum_{\alpha=1}^N J_{k \alpha} (\tau)J_{\ell \alpha}(\tau)
    \label{Q_definition}
\end{equation}
Since their joint distribution $P_\tau(\boldsymbol{s}(\tau), \boldsymbol{h}(\tau))$ is  Gaussian, letting $\langle \cdot\rangle_{\boldsymbol{h}, \boldsymbol{s}}$ denote the average with respect to this measure, we  consider the following covariances between $\boldsymbol{h}(\tau)$ and  $\boldsymbol{s}(\tau)$:
\begin{equation}
R_{i\ell}(\tau) :=\big\langle h_i(\tau) s_\ell(\tau)\big\rangle_{\boldsymbol{h}, \boldsymbol{s}} 
\label{R_definition}
\end{equation}
Using the approach of~\cite{DavidPRL1995, DavidPRE1995}, we  derive, for $\mu_{\rm max} = O(N)$,  the following dynamics for the macroscopic order parameters:
\begin{equation}
\begin{aligned}
    \frac{dQ_{k \ell}}{ d\tau} & = \eta I_{k \ell}  +\eta^2 \tilde{I}_{k \ell} \\ \frac{dR_{i\ell}}{ d\tau} &= \eta \big\langle w_i s_\ell \big\rangle_{\boldsymbol{h}, \boldsymbol{s}}  \\ I_{k\ell} &:=\big\langle h_\ell w_k   + h_k w_\ell \big\rangle_{\boldsymbol{h}, \boldsymbol{s}} \,,\quad   \tilde{I}_{k \ell} :=\big\langle w_k w_\ell\big\rangle_{\boldsymbol{h}, \boldsymbol{s}}  
 \end{aligned}
 \label{QRslow}
\end{equation}
Furthermore, applying Stein's lemma, for $\langle \phi'(h_i)\rangle = \sqrt{2/\pi(1+Q_{ii})}$, we can find $\boldsymbol{T}(\tau)$ from setting the right-hand side of Eq.~\eqref{Tslow1} to zero:
\begin{equation}
\begin{aligned}
    0 =&~ \sigma^2\mathbb{I} + \boldsymbol{A}\boldsymbol{T}(\tau) + \boldsymbol{T}(\tau)\boldsymbol{A}^T \\&+\gamma\left\{{\rm diag}\big(\big\langle\boldsymbol{\phi}'\big\rangle\big) \boldsymbol{R}(\tau)  + \boldsymbol{R}^T(\tau){\rm diag}\big(\big\langle\boldsymbol{
    \phi}'
      \big\rangle\big)\right\} 
     \label{Tslow} \end{aligned} 
\end{equation}
This set of equations~\eqref{QRslow}-\eqref{Tslow} are closed  because  all the averages $I_{k\ell}, \tilde{I}_{k\ell}$ and $\langle w_is_\ell \rangle_{\boldsymbol{h}, \boldsymbol{s}}$  can be expressed in terms of only $\boldsymbol{T}(\tau), \boldsymbol{Q}(\tau)$ and $\boldsymbol{R}(\tau)$, see Appendix \ref{sec:formulation} for details.

In summary,  with $\tau$ being the time on which the signal varies, our macroscopic theory characterizes  an ensemble of cells of a given type  in terms of the correlations among target genes $\boldsymbol{T}(\tau)$, tracks the effectors' state in terms of their correlations $\boldsymbol{Q}(\tau)$   and quantifies the  covariances $\boldsymbol{R}(\tau)$ between these components of  the intracellular signaling
pathway. The  behavior of the original system in Eqs.~\eqref{learning_dynamics}-\eqref{fast_dynamics_tau} can now be fully described by $\boldsymbol{T}(\tau)$, $\boldsymbol{Q}(\tau)$ and $\boldsymbol{R}(\tau)$. The  evolution of these quantities can be numerically computed using Eqs.~\eqref{QRslow}-\eqref{Tslow}. Their steady-state solutions  correspond  to  distinct time-invariant covariations   of the intracellular components in the presence of time-varying signal. Therefore, from now on, to understand these distinct regimes of cellular coordination, we shall  focus on these macroscopic covariances. 

\section{Results}
\label{sec:results}


\begin{figure*}[t]
    \centering
\includegraphics[width=0.32\linewidth]{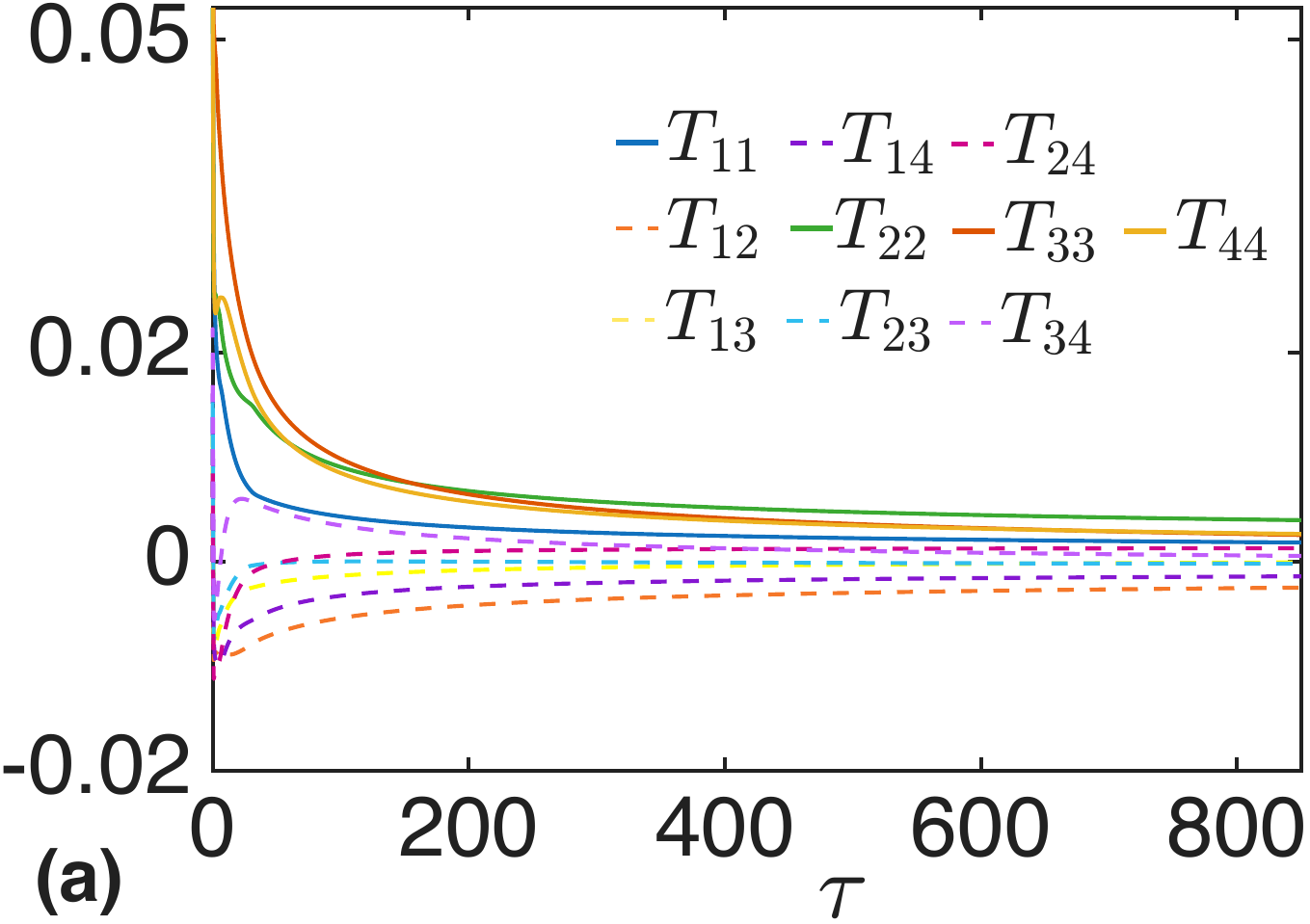}
\includegraphics[width=0.32\linewidth]{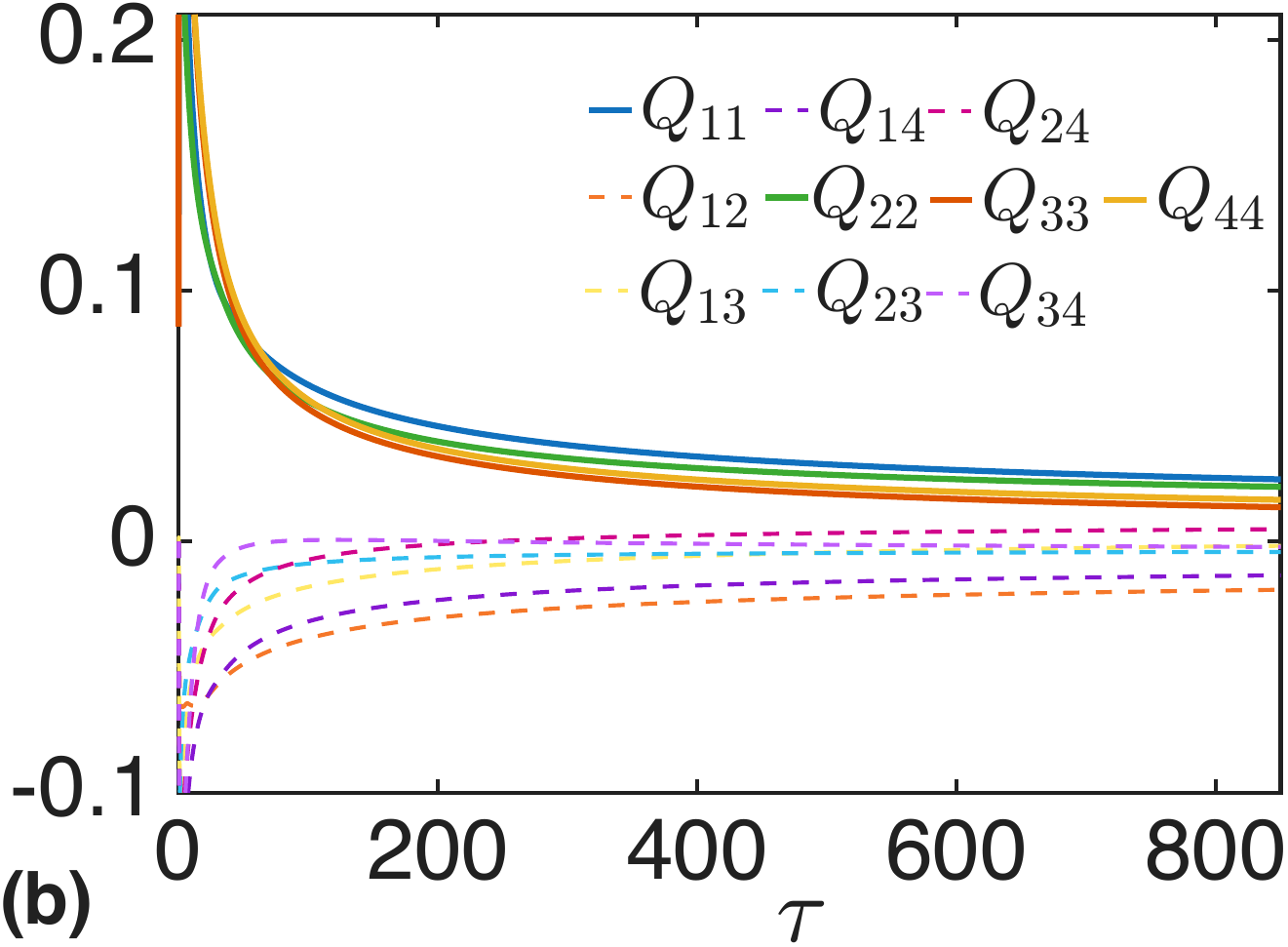}
\includegraphics[width=0.32\linewidth]{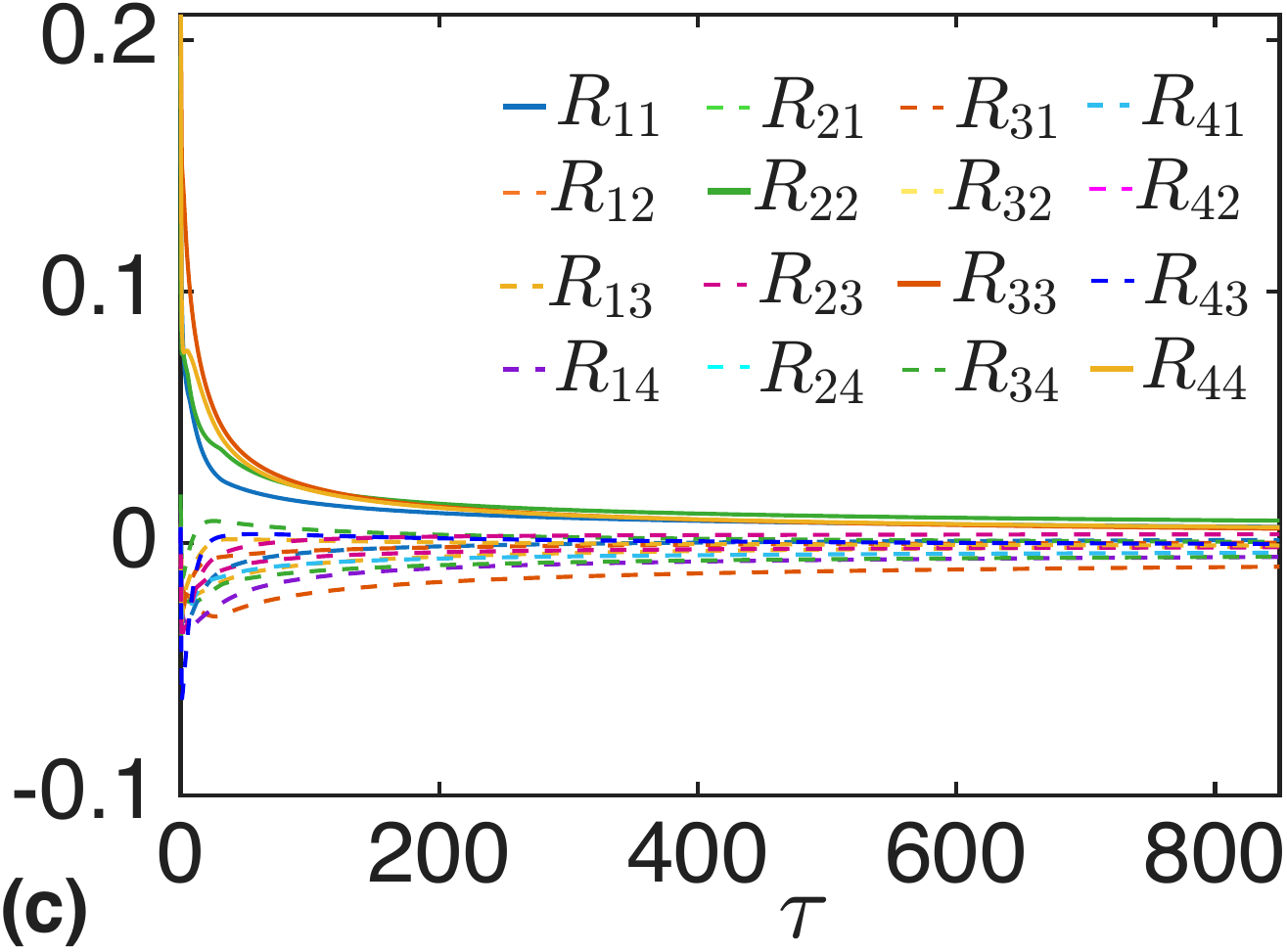}
\vskip0.2cm
\includegraphics[width=0.32\linewidth]{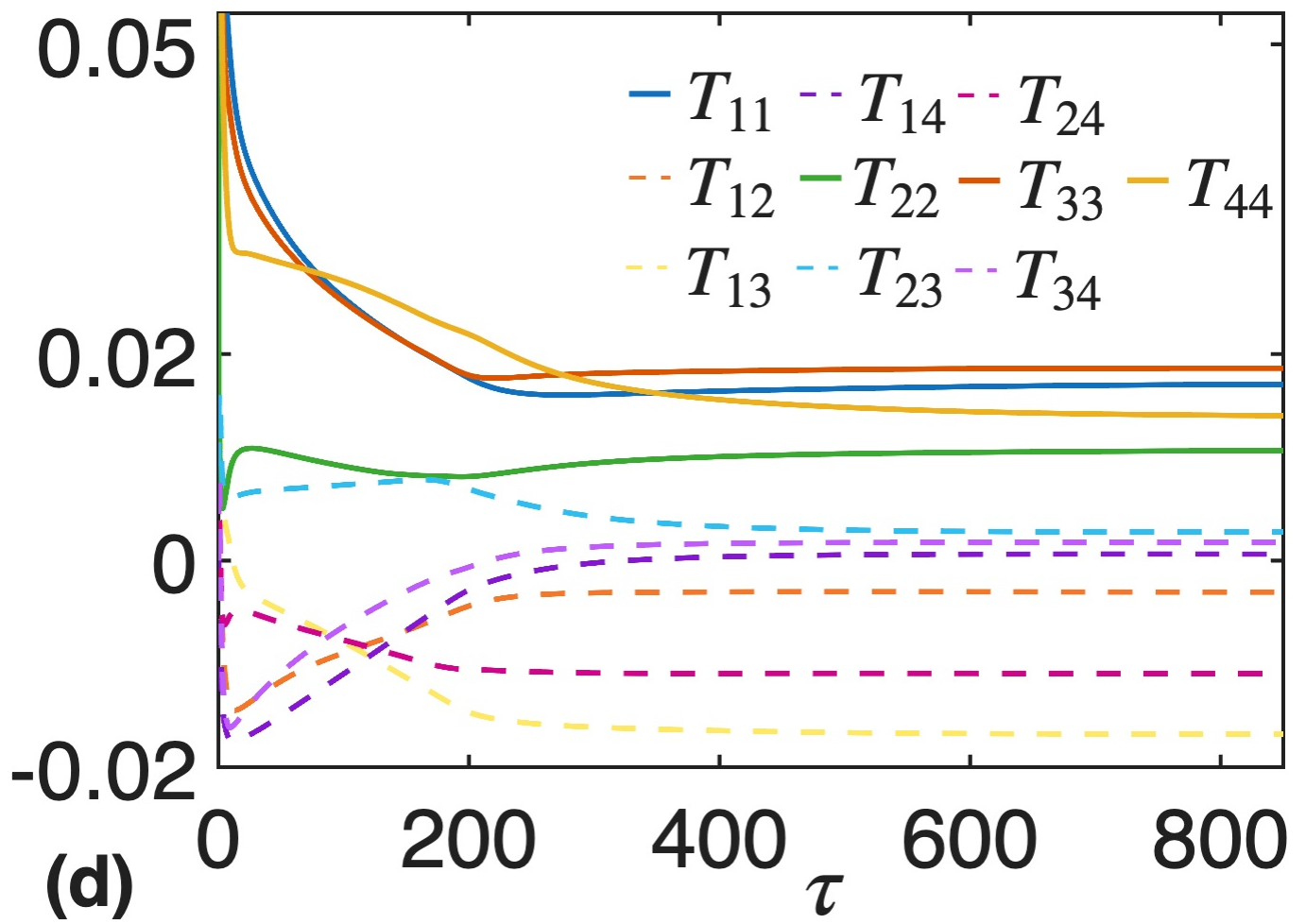}
\includegraphics[width=0.32\linewidth]{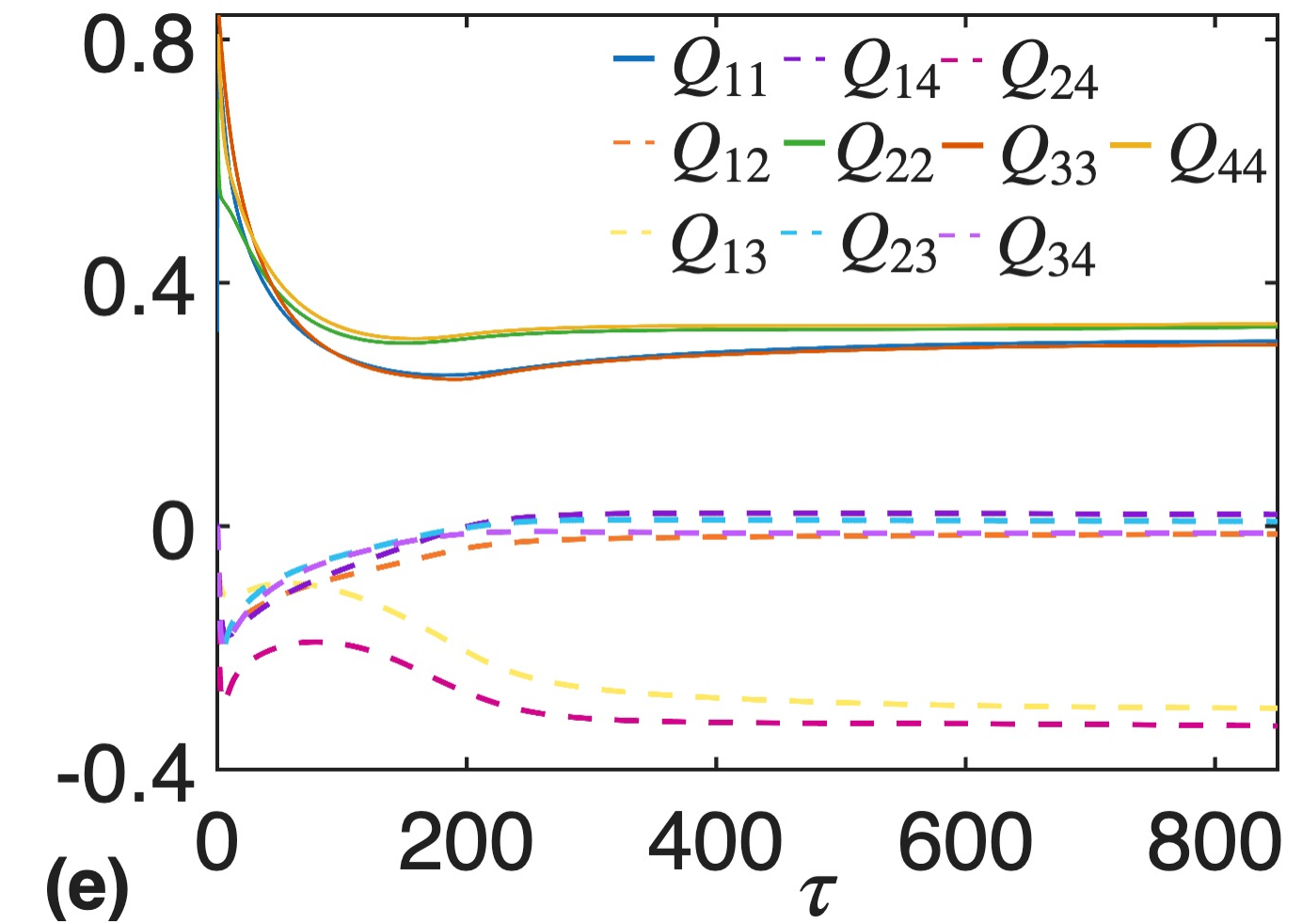}
\includegraphics[width=0.32\linewidth]{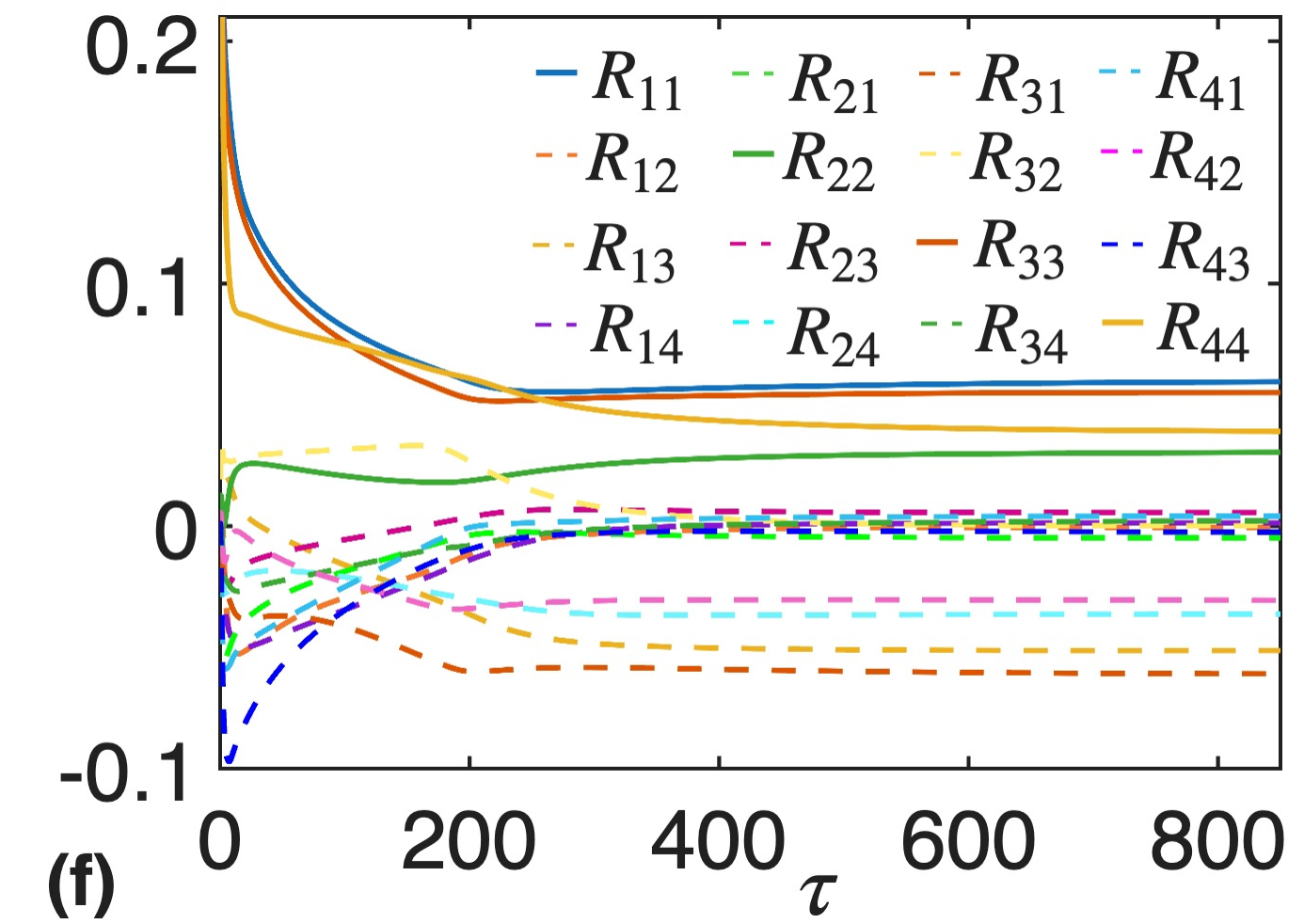}
\vskip0.2cm
\includegraphics[width=0.32\linewidth]{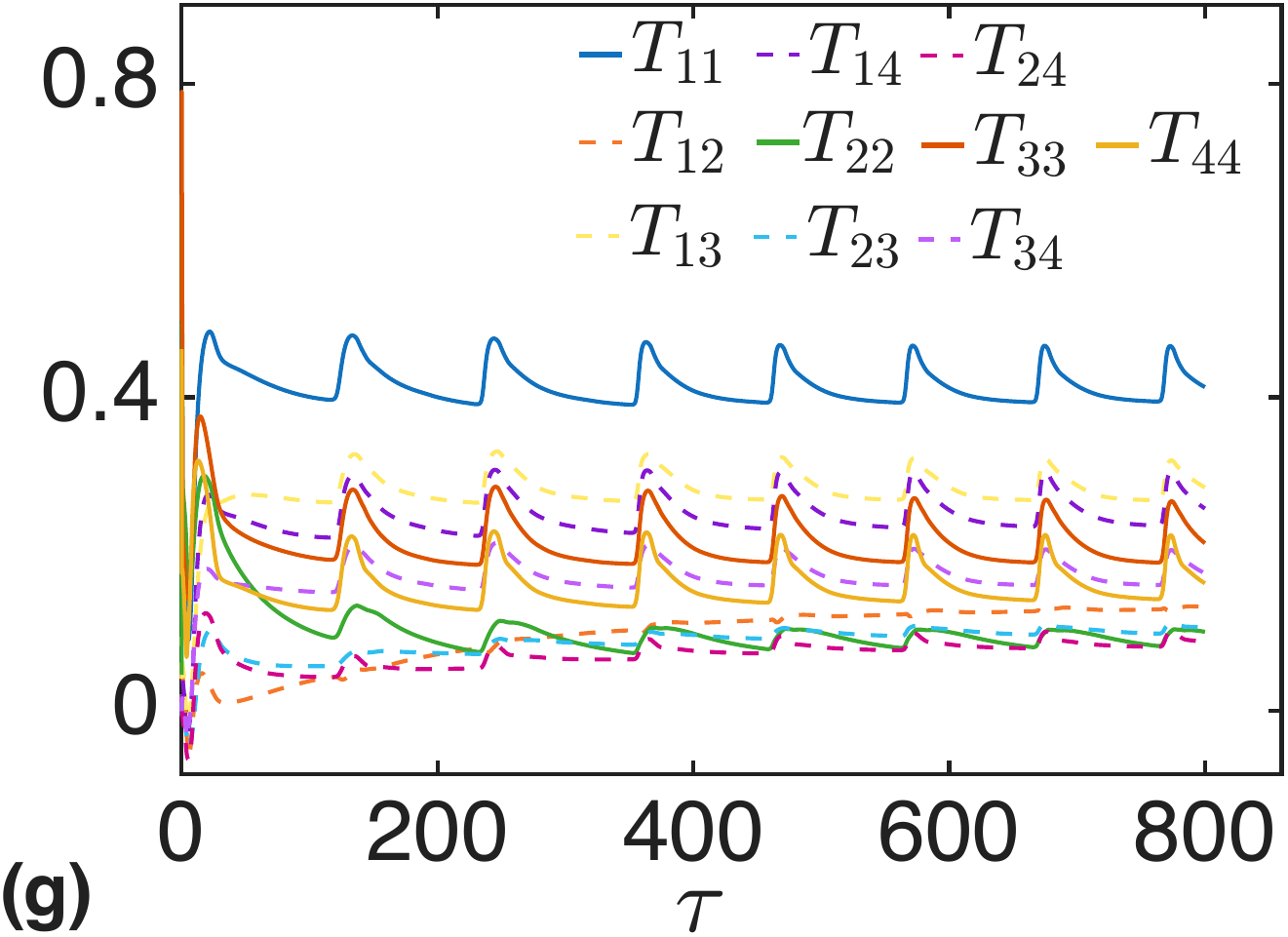}
\includegraphics[width=0.32\linewidth]{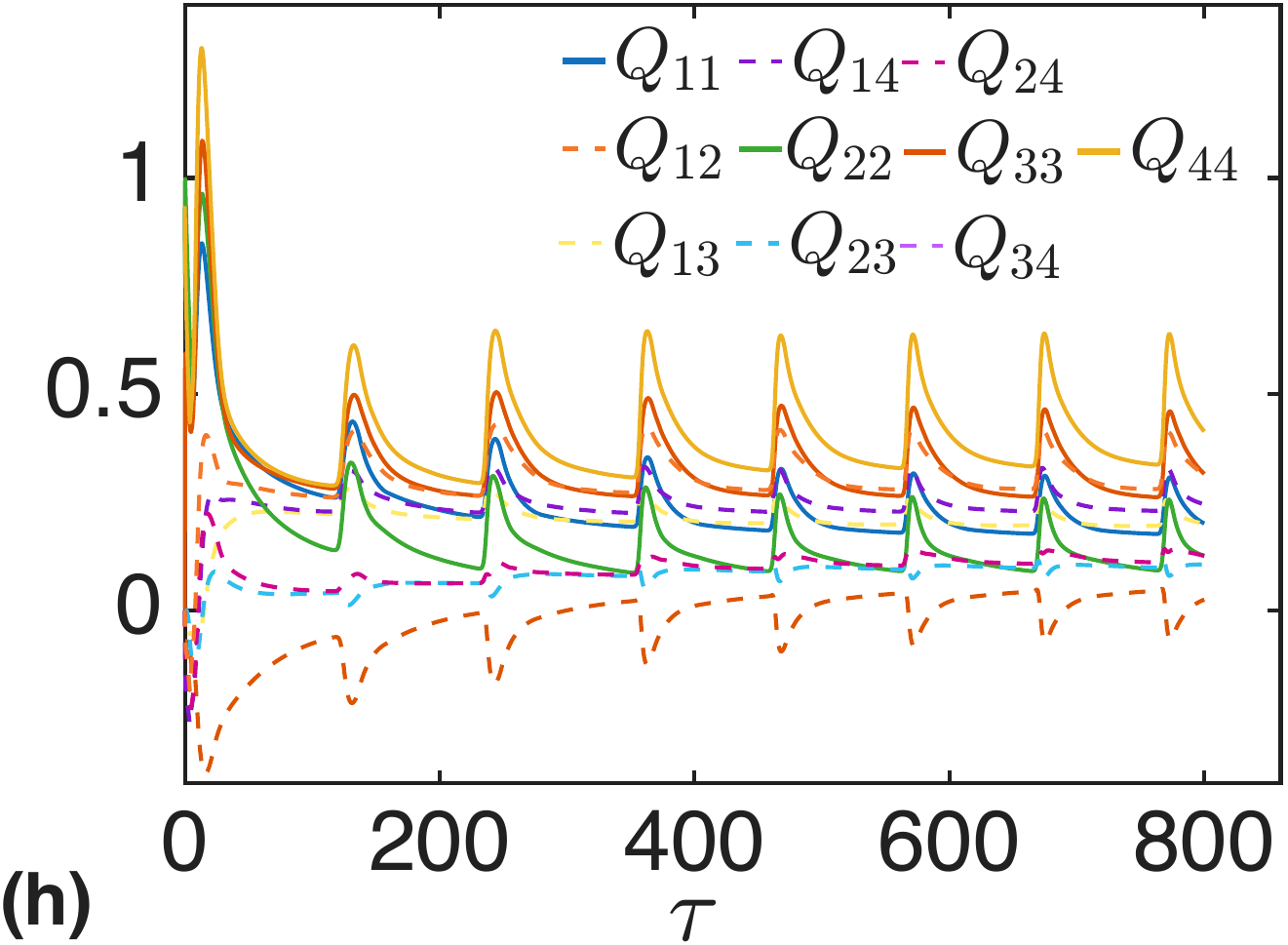}
\includegraphics[width=0.32\linewidth]{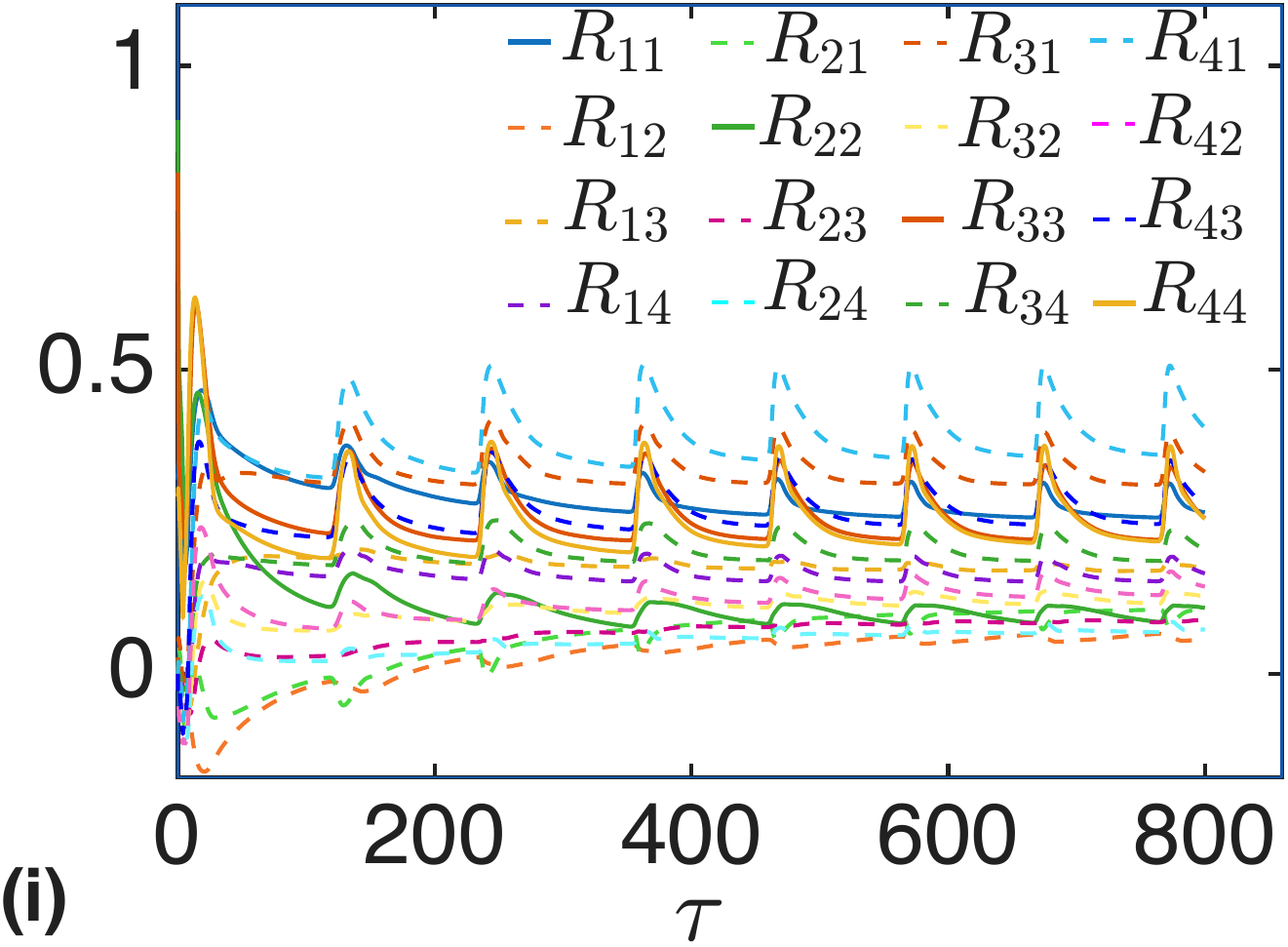}
\caption{The target gene correlations $\boldsymbol{T}$, the effectors' covariations $\boldsymbol{Q}$,  the cross-correlations between the target genes and their effectors $\boldsymbol{R}$, all  as function of the slow time $\tau$ for $(\eta, \gamma) =( 0.5,0.5)$ in \textbf{(a)}-\textbf{(c)} $(\eta, \gamma) =( 0.9,0.5)$ in \textbf{(d)}-\textbf{(f)}  for  $(\eta, \gamma) =( 0.9,0.9)$ in \textbf{(g)}-\textbf{(i)}.  Here     $c = 0.2$, $\sigma = 0.01$, $K=4$ and $\boldsymbol{W}$ is a random \emph{asymmetric} matrix of i.i.d.   elements drawn from $\mathcal{U}(0,1)$ -- the uniform distribution in the interval $(0,1)$. Initial conditions for $\boldsymbol{Q}$ and $\boldsymbol{T}$ are specified by $Q_{ii}$ and $T_{ii}$, both  drawn from   $\mathcal{U}(0,1)$,  while $Q_{ij}$ and $T_{ij}$ -- are sampled from  the uniform distribution  $10^{-6}\times\mathcal{U}(0,1)$. All elements of $\boldsymbol{R}(\tau =0)$ are drawn from the uniform distribution $10^{-6}\times\mathcal{U}(0,1)$.}
\label{fig2}
\end{figure*}


Within the framework introduced in the previous section, a  cellular system robust to varying signal is  one that exhibits a non-trivial  long-time correlations $\big(\boldsymbol{T}(\tau), \boldsymbol{Q}(\tau),\boldsymbol{R}(\tau)\big)$. We are particularly interested in parameter values for which such robustness  can emerge even for signal fluctuating strongly enough to overcome the activation threshold and induce cell-type switching. 
To this end, we first check whether our system~\eqref{QRslow}-\eqref{Tslow}  can relax to a  fixed point $(\boldsymbol{Q}_*, \boldsymbol{R}_*, \boldsymbol{T}_*)$  for a wide range of parameters and  interaction matrix $\mathbf{W}$ such that $\boldsymbol{A}:=-\mathbb{I}+ c\boldsymbol{W}$ is a Hurwitz stable matrix. 
Such an attractor corresponds to a given cell type. In the following, we chose $K=4$ to match  the real system  of four target genes in \cite{Pezzotta}.  However,  it is easily scaled up to capture  larger systems. 

Since we aim at capturing the generic behavior of GRNs, instead of using  GRNs inferred from gene expression data, 
we consider $\boldsymbol{W}$ as a random \emph{asymmetric} matrix, whose entries are independently and identically distributed (i.i.d.) variables drawn from $\mathcal{U}(0,1)$,  the uniform distribution in $(0,1)$ to remove one microscopic source of negative correlations: direct negative entries in $\boldsymbol{W}$.  In this regard, unlike the  work of \cite{BALASKAS}, where Gli's adaptation to morphogen variation was explained by  repressive interactions, here our setup does not assume any repression between target genes.  The network is  fully-connected and without self-loops, unless stated otherwise. For brevity, frow now on,  we will refer $T_{i\neq j}$ to as $T_{ij}$,  $Q_{i\neq j}$ to as $Q_{ij}$ and  $R_{i\neq j}$ to as $R_{ij}$.

By varying $(\eta, \gamma)$ continuously at   fixed $(c,
\sigma) = (0.2, 0.01)$, we can scrutinize how $\eta$  and $\gamma$ modulate the cell type $(\boldsymbol{Q}_*, \boldsymbol{R}_*, \boldsymbol{T}_*)$, where $\eta$ tunes the rate of change of the effectors' dynamics and $\gamma$ regulates its impact on the target genes' dynamics. We obtain three different behaviors, that are illustrated  by the steady-state solutions of Eqs.~\eqref{QRslow}-\eqref{Tslow} at   $(\eta, \gamma) =(0.5, 0.5)$, $(\eta,\gamma) = (0.9, 0.5)$, and $(\eta,\gamma) = (0.9, 0.9)$ in Fig.~\ref{fig2} \textbf{(a)}-\textbf{(c)},   \textbf{(d)}-\textbf{(f)}, and   \textbf{(g)}-\textbf{(i)}, respectively. In the first case, $T_{ii}$ approaches $O(\sigma^2/2)$
and $T_{ij}$  decays to zero at large $\tau$, so that both are vanishingly small in the long time limit; in the   second case, $T_{ii}$ remains well-separated from $T_{ij}$, implying  an increased rate of adaptation results  in  significantly distinct target vectors which the fast dynamics  converge to;
in the last case, we observe oscillatory solution when the controlling impact  of the effector, $\gamma$, and  the adaptation strength, $\eta$,  are both strong. 

Next, 
we find a similar behavior of the overlap   $\boldsymbol{Q}$, namely,  a vanishing gap between $Q_{ii}$
and $Q_{i j}$  for $\eta = 0.5$ in Fig.~\ref{fig2} \textbf{(b)}, where both become less relevant, and a finite gap for $\eta = 0.9$ in Fig.~\ref{fig2} \textbf{(e)}, where they follow their respective target vectors   $\boldsymbol{T}$ to achieve separation ($Q_{ii} \gg Q_{ij}$); and Fig.~\ref{fig2} \textbf{(h)}, where $Q_{ii}$ and  $Q_{ij}$, both oscillate. Note that, from Eq.~\eqref{Q_definition}, a solution with $Q_{ii}\rightarrow 0$ does not have any interesting interpretation as it corresponds to $J_{i \alpha}(\tau\rightarrow \infty)  = 0$, $\forall\, \alpha$, implying that morphogen has no effect on the effector. In this regard, with weak adaptation rate $\eta=0.5$, a single cell can only remain robust against a varying signal if  it is uncoupled from the latter.

The convergence of $\boldsymbol{T}$ and $\boldsymbol{Q}$  to non-zero values $(\boldsymbol{T}_*,\boldsymbol{Q}_*)$ is accompanied by the establishment of a stable  covariance   $\boldsymbol{R}_*$ between the regulators and their  effectors in   Fig.~\ref{fig2} \textbf{(f)} for $(\eta, \gamma)= (0.9,0.5)$. Here a  high separation between $R_{ii}$ and $R_{ij}$ is observed in the long time limit (compared to the case of $\eta=0.5$ in Fig.~\ref{fig2} \textbf{(c)}, where $R_{ii}$ and $R_{ij}$, both decays to zero). 
Taken altogether, Figs.~\ref{fig2} \textbf{(a)}-\textbf{(f)}  shows that,  strong feedback can indeed stabilize  the interlinked dynamics,  maintaining a time-invariant correlation pattern in  the present of morphogen temporal variations. This confirms our hypothesis that the level of  fluctuations in gene expression  is endogenously controlled by the receiving cell  through self-regulation. Thanks to the latter, target genes  can indeed determine their own attractors by tuning  their effectors accordingly and in turn, these effectors can achieve a level of activity consistent with what determined by the attractor of the GRN dynamics. This consistency between these two  modules constitutes our ``cooperative'' mechanism of adaptation. 

 Finally, let us remark on the observed  oscillating solution at high driving rate $\gamma$  and high adaptation rate $\eta$  in Fig.~\ref{fig2} \textbf{(g)}-\textbf{(i)}. Given  $\boldsymbol{A}:=-\mathbb{I}+ c\boldsymbol{W}$ is  Hurwitz stable,   at a fixed effectors' state, the  linearized fast GRN dynamics  always converges to a stationary state. Therefore,   the relaxational-oscillatory behavior of $(\boldsymbol{T},\boldsymbol{Q}, \boldsymbol{R})$ on the slow time $\tau$ can only be attributed to an under damped  feedback loop between $(\boldsymbol{Q}, \boldsymbol{R})$ and $\boldsymbol{T}$. Specifically, 
nonlinear activations make the system drift slowly along one quasi-stable branch, then switch rapidly when the  feedback crosses a threshold.  
Thus, this oscillatory regime  plausibly corresponds to 
a dynamic  adaptation mode.

\begin{figure}
    \centering
\includegraphics[width=0.47\linewidth]{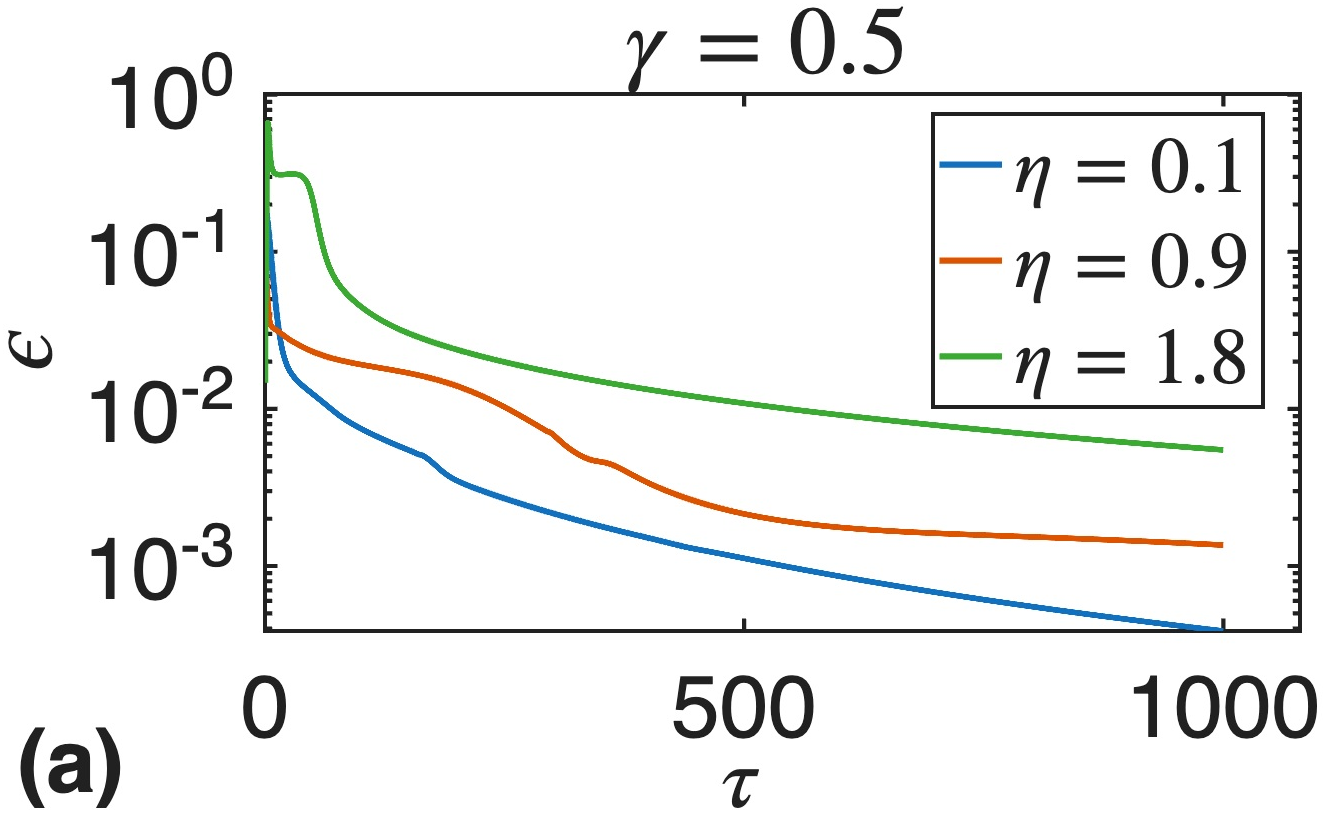}
\includegraphics[width=0.47\linewidth]{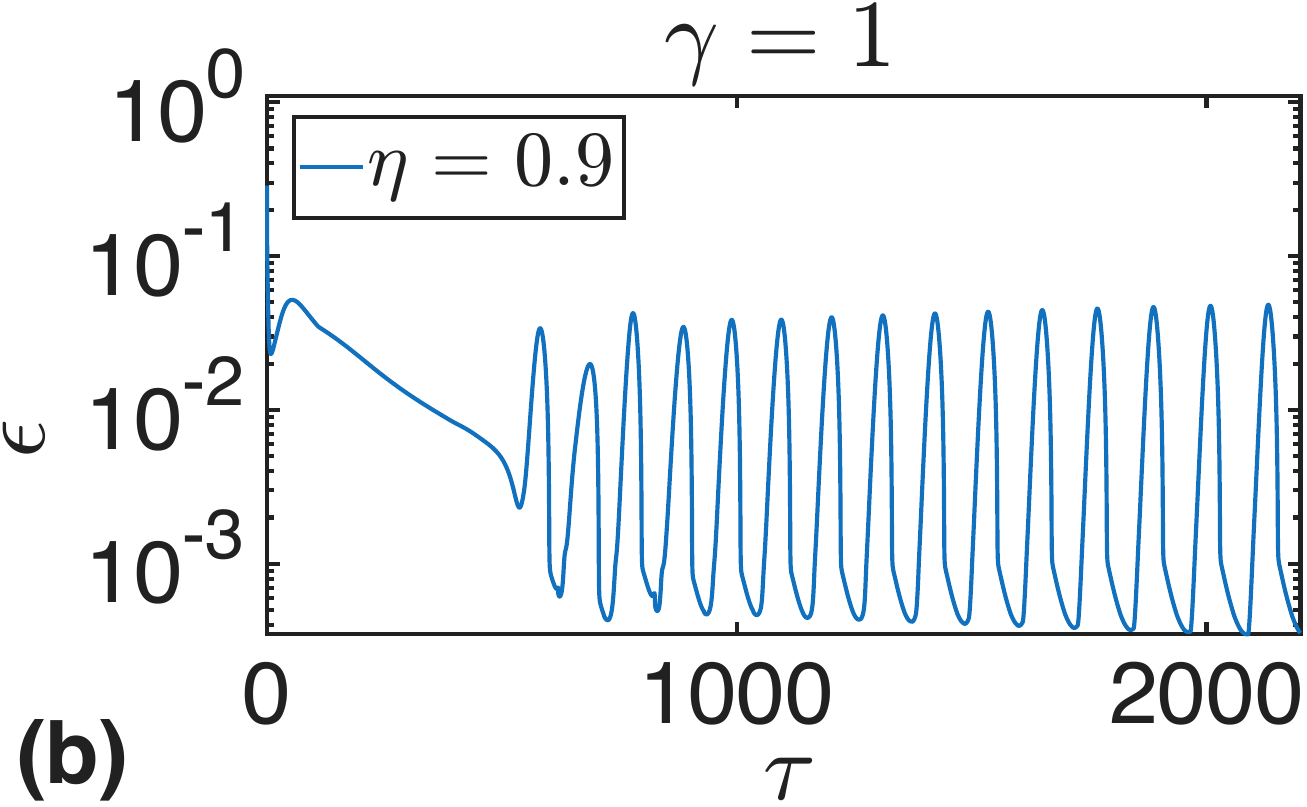}
\caption{$\gamma=0.5$ \textbf{(a)}  and $\gamma = 1$ \textbf{(b)}. 
Here    $c = 0.2$, $\sigma = 0.01$, $K=4$, and $\boldsymbol{W}$ is an \emph{asymmetric} matrix of uncorrelated elements drawn from $\mathcal{U}(0,1)$ -- the uniform distribution in the interval $(0,1)$. Initial conditions for $\boldsymbol{Q}$ and $\boldsymbol{T}$ are specified by $Q_{ii}$ and $T_{ii}$, both  drawn from   $\mathcal{U}(0,1)$,  while $Q_{ij}$ and $T_{ij}$ -- are sampled from  the uniform distribution $10^{-6}\times\mathcal{U}(0,1)$. All elements of $\boldsymbol{R}(\tau =0)$ are drawn from the uniform distribution $10^{-6}\times\mathcal{U}(0,1)$.}
\label{fig3}
\end{figure}

Next we examine how well such consistency is satisfied  by tracking  the development of $\epsilon(\tau)$   for $\gamma = 0.5$ in Fig.~\ref{fig3}\textbf{(a)}.  
In both cases, $\epsilon$  relaxes towards zero, with a convergence rate depending on $\eta$. Counterintuitively, the smaller $\eta$, the faster the decay of $\epsilon$ and the lower the value it attained. 
This can possibly be explained by the impact of the larger update steps on the $\boldsymbol{W}$ dynamics, rapidly changing the target correlations and thus making the convergence towards lower $\epsilon$ values more challenging.
Note the existence of a plateau-like region during the transient dynamics of $\epsilon$ at high $\eta=1.8$. This might correspond to the well-known  symmetry-breaking phenomenon in online training, which exhibits a symmetric phase before the learning process converges to a symmetry-broken solution~\cite{DavidPRE1995}. It is known that the adaptation rate $\eta$ acts as a bifurcation parameter, leading to  various bifurcations as well as multistability even for a fixed target vector $\boldsymbol{T}$ \cite{Biehl_1996}. Interestingly, Fig.~\ref{fig3} \textbf{(b)} illustrates an oscillation of the mismatch that occurs at  $\gamma=1$  and $\eta =0.9$, since all the covariances $(\boldsymbol{T}, \boldsymbol{Q}, \boldsymbol{R})$ oscillate in this case. Intuitively speaking, the effectors and target genes try to match each other, but strong feedback causes them to repeatedly overshoot. With the sigmoidal nonlinearity $\phi$, this overshooting can become pulse-like as seen in  Fig.~\ref{fig2} \textbf{(g)}-\textbf{(i)}. 

\textbf{Symmetric solution.} In Fig.~\ref{fig2}, results are obtained by time-integrating the full fast-slow coupled dynamics for $(\boldsymbol{T}, \boldsymbol{Q}, \boldsymbol{R})$ with a single random interaction  matrix $\boldsymbol{W}$.  To make an analytical progress on understanding the model's different regimes reported in the previous section, we assume self-averaging, so that we can consider the typical behavior of an ensemble of systems Eqs.~\eqref{QRslow}-\eqref{Tslow}, each has its own matrix $\boldsymbol{W}$.  For $\eta \ll 1$, neglecting the term proportional to $\eta^2$, we can look for a steady-state symmetric solution   $Q_{ii} =Q_{ij}=q$; $R_{ii} = R_{ij}= r$;  $T_{ii}= u$ and $T_{ij}=v$
to the  set of equations~\eqref{QRslow}-\eqref{Tslow}, as detailed in Appendix~\ref{sec:derivation_sym}. Using this ansatz we obtain the following relationship between $q$ and $r$:
\begin{equation}
\begin{aligned}
    r&= \pm \kappa q
    \\ \kappa^2  &:=1+\frac{\sigma^2}{2\mu}\left[\frac{1-cw(K-2)}{1+cw}-\frac{1}{K}\right] \\ 
    \end{aligned}
    \label{symmetric_with_noise}
\end{equation}
 where  $w:= \langle W_{ij}\rangle_{\boldsymbol{W}}$. As detailed in Appendix \ref{sec:derivation_sym}, $q$ is determined from a positive root of the equation:
 \begin{equation}
 \begin{aligned}
     q
&=
\frac{\sigma^2}{2K}\sqrt{\frac{\pi}{2}}~\left[
 \gamma_c \kappa^2- \frac{\gamma\kappa}{\sqrt{1+q}}
\right]^{-1}\\
  \gamma_c &:= \sqrt{\frac{\pi}{2}}~\big[1-cw(K-1)\big]\,,
  \end{aligned}
  \label{q_symmetric_with_noise}
 \end{equation}
 while the target gene correlations are given by
\begin{equation}
    u =\kappa^2 (1+q)-1 \,,\quad  v=u - \frac{\sigma^2}{2(1+cw)}\,.
    \label{t_symmetric_with_noise}
\end{equation}
Note that we require $1-cw(K-1) >0$ for $\langle \boldsymbol{A}\rangle_{\boldsymbol{W}}$ being Hurwitz stable~\footnote{We emphasize that replacing $\boldsymbol{W}$ by its mean is a strong approximation and it is not exactly equivalent to the self-averaging assumption}.  
As $\sigma \rightarrow 0$, we have $\kappa\rightarrow 1$, then from Eq.~\eqref{q_symmetric_with_noise} we obtain the following explicit solution 
\begin{equation}
   u=v=q=\pm r= \left \{ \begin{array}{l} \displaystyle 0  \,,\qquad\qquad \quad  \gamma\leq \gamma_c\\
\displaystyle   \left(\frac{\gamma}{\gamma_c}\right)^2 -1\,,\,\,\,\, \gamma> \gamma_c
\end{array} \right. 
\label{symmetric_no_noise}
\end{equation}
Here we identified $\gamma_c$ as a critical threshold  separating  a trivial
symmetric state  $(q,r,u,v)=(0,0,0,0)$ from a nontrivial symmetric one given by Eq.~\eqref{symmetric_no_noise}.

   Before moving on, let us remark that  solution~\eqref{symmetric_with_noise} shows a linear scaling relationships between $q$ and $r$. This is in marked difference from the classical online-learning framework~\cite{DavidPRE1995}, where, as the overlap matrix $\boldsymbol{T}$ is fixed, $r\propto q^{1/2}$ according to a geometrical picture that elements $R_{ij}$ of $\boldsymbol{R}$
are just projections of the interpretation matrix  $\boldsymbol{J}$ onto the $K$-dimensional  subspace spanned by the basis  vectors $\boldsymbol{e}_n$, $n=1,\cdots, K$ with $( \boldsymbol{e}_n, \boldsymbol{e}_m)= T_{nm}$.  Naturally, the dynamically changing length of the subspace vectors requires normalization in order to see this effect more clearly. 

\begin{figure}
    \centering
\includegraphics[width=0.96\linewidth]{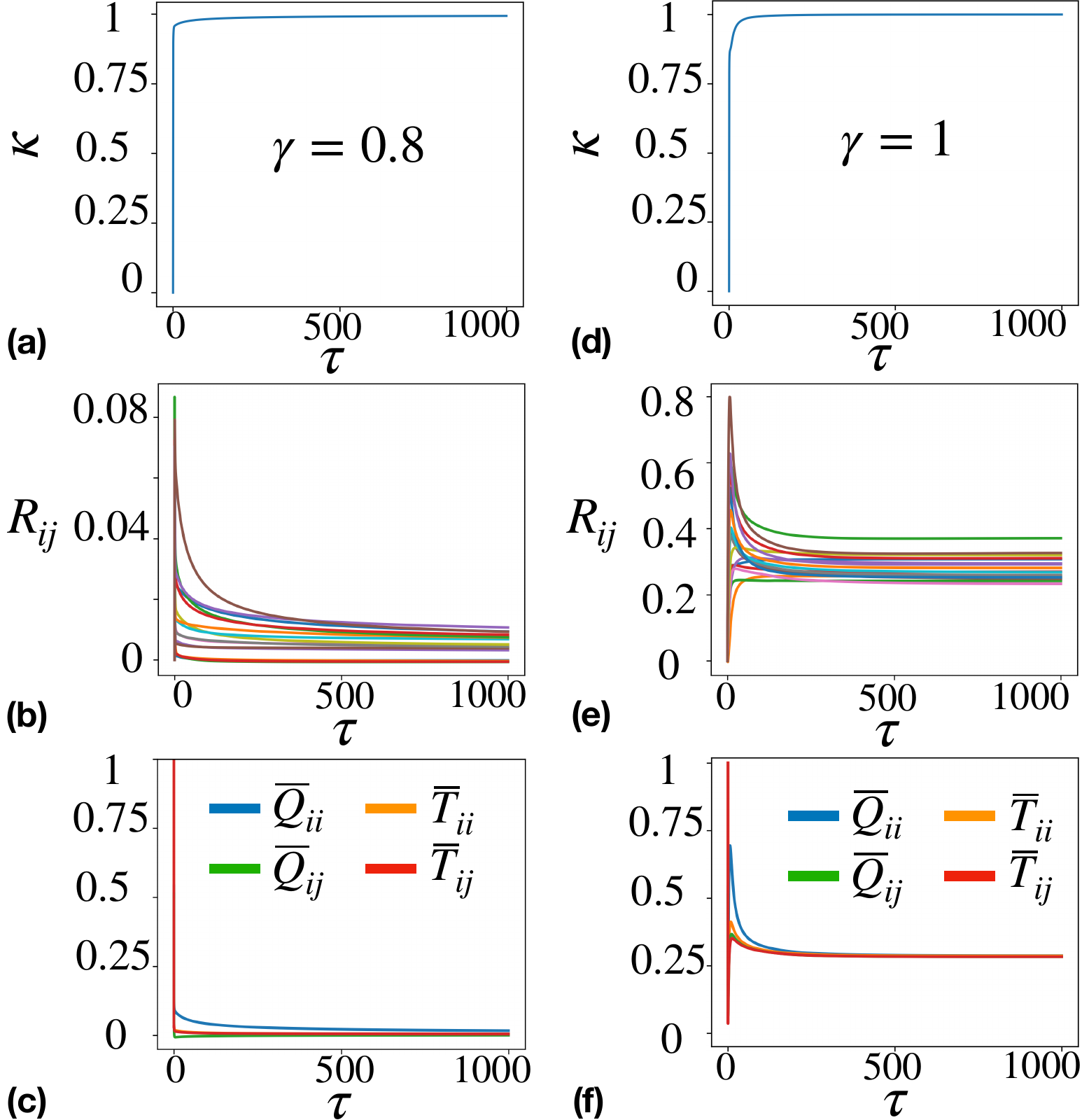} 
\caption{\textbf{Symmetric solution} for   $\gamma =0.8$ (\textbf{(a)}-\textbf{(c)})  and  $\gamma =1$ (\textbf{(d)}-\textbf{(e)}). From top to bottom: the ratio   $\kappa = \bar{r}/\bar{q}$ with $\bar{r}\equiv K^{-2}\Big[\sum_{i} R_{ii} +\sum_{i\neq j} R_{ij}\Big]$ and $\bar{q}\equiv K^{-2} \Big[\sum_{i} Q_{ii} +\sum_{i\neq j} Q_{ij}$\Big], all covariances $R_{ii}$ and $R_{ij}$, the averaged diagonal  and off-diagonal  overlaps $(\overline{Q}_{ii}, \overline{T}_{ii})$ and $(\overline{Q}_{ij}, \overline{T}_{ij})$, with $\overline{Q}_{ii}:=K^{-1}\sum_{i=1}^K Q_{ii}$ and   $\overline{Q}_{ij}:=2 \sum_{i<j}Q_{ij}/[K(K-1)]$. 
Here  $\eta =0.5$,   $c = 0.2$, $\sigma = 0.01$, $K=4$, and $\boldsymbol{W}$ is an \emph{asymmetric} matrix of uncorrelated elements drawn from $\mathcal{U}(0,1)$ -- the uniform distribution in $(0,1)$. We use symmetric initial conditions:  $Q_{ii} =Q_{ij}= 0.5$, $R_{ii}=R_{ij} =10^{-6}$, $T_{ii} = 1+\sigma^2/(2*(1+c \langle W\rangle))$, $T_{ij} = 1$.}
\label{fig4}
\end{figure}

\begin{figure}
    \centering
\includegraphics[width=0.45\linewidth]{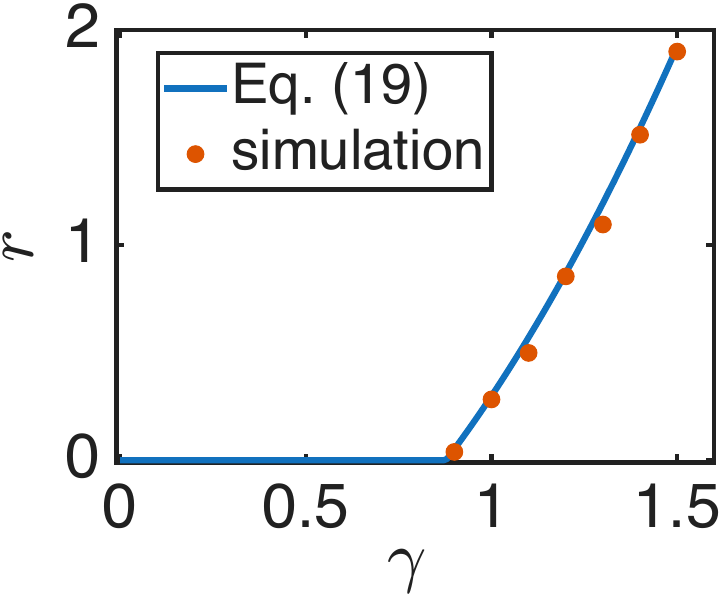} 
\caption{Symmetric solution as function of  $\gamma$. We compare analytical value of $r$ given by Eq.~\eqref{symmetric_no_noise} (blue line) with the value  with $\langle \bar{r}\rangle_W$ obtained from direct  time-integration of Eqs.~\eqref{QRslow}-\eqref{Tslow} (red dots), where $\bar{r}\equiv K^{-2}\Big[\sum_{i} R_{ii} +\sum_{i\neq j} R_{ij}\Big]$.
Here the average is taken over an ensemble of 100 \emph{asymmetric} matrices $\boldsymbol{W}$  of uncorrelated elements drawn from $\mathcal{U}(0,1)$,  error bars are of the same size as the symbols and hence not shown. $\eta =0.5$,   $c = 0.2$, $\sigma = 0.01$, $K=4$. We use symmetric initial conditions:  $Q_{ii}=Q_{ij}= 0.5$, $R_{ii}=R_{ij} =10^{-6}$, $T_{ii} = 1+\sigma^2/(2*(1+c \langle W\rangle))$, $T_{ij} = 1$.}
\label{fig5}
\end{figure}

In Fig.~\ref{fig4}, we use direct integration of Eqs.~\eqref{QRslow}-\eqref{Tslow} with symmetric initialization at  $\eta=0.5$ and  $\sigma=0.01$ to  verify the prediction of Eq.~\eqref{symmetric_with_noise}. Specifically, using the  values of $\boldsymbol{R}$ and $\boldsymbol{Q}$ obtained for  a single random interaction matrix $\boldsymbol{W}$, we plot
$\kappa = \bar{r}/\bar{q}$ as function of time $\tau$ with $\bar{r}\equiv K^{-2}\Big[\sum_{i} R_{ii} +\sum_{i\neq j} R_{ij}\Big]$ and $\bar{q}\equiv K^{-2} \Big[\sum_{i} Q_{ii} +\sum_{i\neq j} Q_{ij}$\Big]. We observe a fast  convergence of $\kappa=\bar{r}/\bar{q}$
towards  values close to 1 as the system relaxes to steady state, even though each of the individual elements of $\boldsymbol{R}$ and $\boldsymbol{Q}$ converges to their own values. This confirms the asymptotics $\kappa \rightarrow 1$ as $\sigma\rightarrow 0$ of Eq.~\eqref{symmetric_with_noise}. Moreover,  Fig.~\ref{fig5} demonstrates the existence of a threshold $\gamma_c\simeq 0.88$, below which only trivial symmetric solution is realized, but beyond which we find a positive symmetric solution that grows with $\gamma$. Finally, we find excellent agreement between the analytical prediction of  Eq.~\eqref{symmetric_no_noise} and numerical results obtained  by averaging over an ensemble of independent steady states, each is obtained by time-integration of Eqs.~\eqref{QRslow}-\eqref{Tslow} under a given $\boldsymbol{W}$.

At  a symmetric fixed point given by Eq.~\eqref{symmetric_with_noise}, one can show that the Jacobian 
 can be decomposed into the two invariant sectors: a
symmetry-preserving (longitudinal) sector, and a symmetry-breaking (transverse) one. A linear stability analysis can then be carried out to identify the onset of an instability 
  due to  perturbation along the latter sector \footnote{Longitudinal perturbation does not push the solution leaving the symmetric manifold, so is not responsible for specialization}. Since the Jacobian's expression is unwieldy, we relegate it in  Appendix \ref{sec:Jacobian}.  In integration of Eqs.~\eqref{QRslow}-\eqref{Tslow}, such transversal perturbations can be directly implemented as  asymmetric initializations.



\begin{figure}[t]
    \centering  
\includegraphics[width=0.95\linewidth]{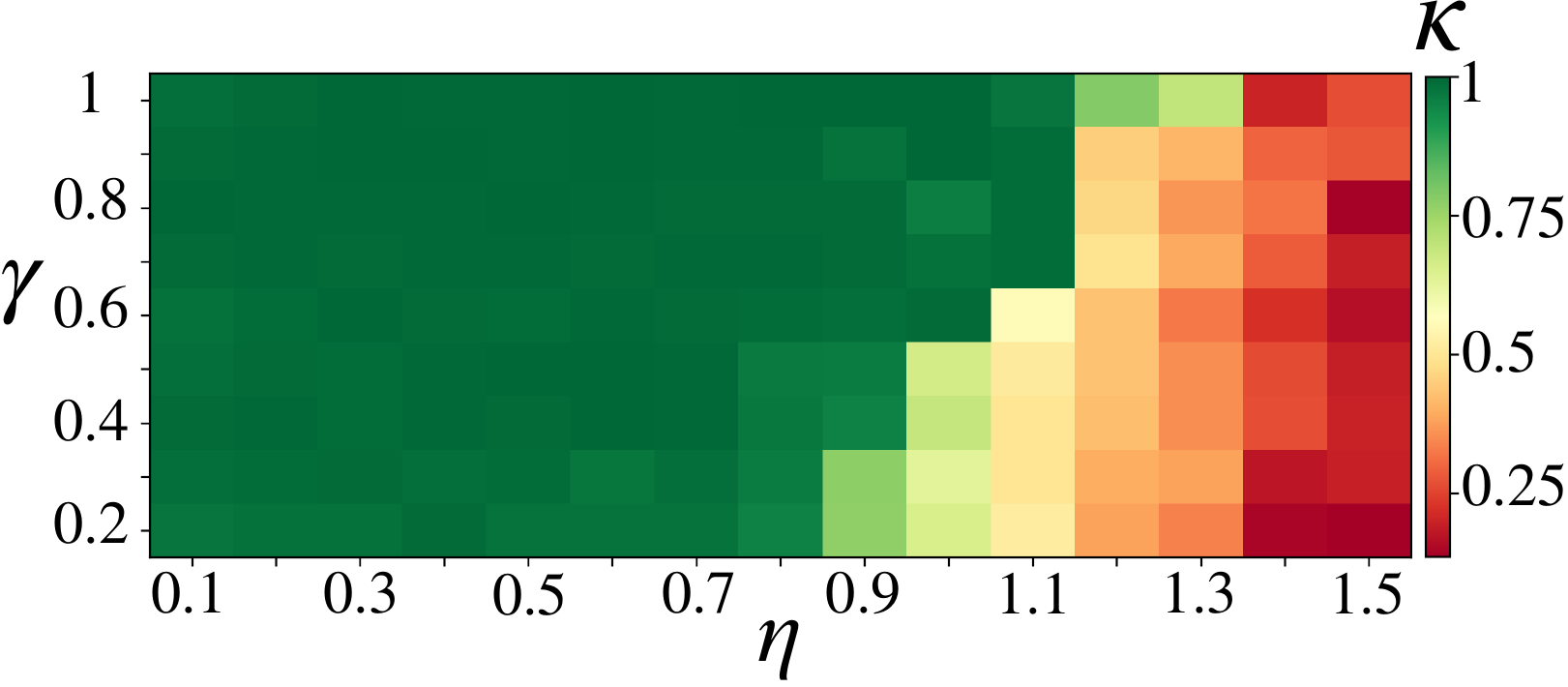}
\caption{\textbf{Phase diagram} 
for  $c = 0.2$, $\sigma = 0.01$, $K=4$. Results are obtained by averaging over an ensemble of 100 \emph{asymmetric} matrices $\boldsymbol{W}$  of uncorrelated elements drawn from $\mathcal{U}(0,1)$. Asymmetric initial condition for $\boldsymbol{Q},\boldsymbol{T},\boldsymbol{R}$ as explained in Fig~\ref{fig2}.}
    \label{fig6}
\end{figure}

\begin{figure}
    \centering  
\includegraphics[width=0.78\linewidth]{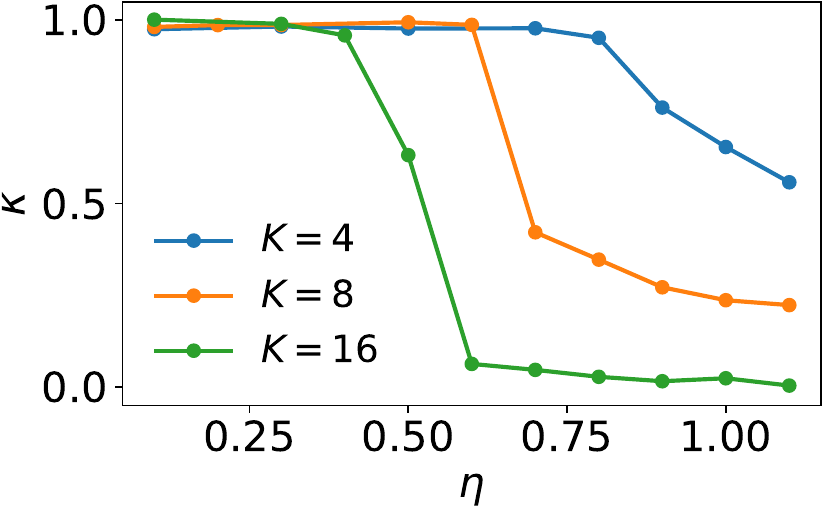}
\caption{$K$-dependence of the transition for $\gamma=c = 0.2$, $\sigma = 0.01$. Results are obtained by averaging over an ensemble of 100 \emph{asymmetric} matrices $\boldsymbol{W}$  of uncorrelated elements drawn from $\mathcal{U}(0,1)$, error bars are of the same size as the symbols and hence not shown. Asymmetric initial condition for $\boldsymbol{Q},\boldsymbol{T},\boldsymbol{R}$ as explained in Fig~\ref{fig2}.}
    \label{fig7}
\end{figure}

Figure~\ref{fig6} presents the steady-state phase diagram in terms of $\kappa$ obtained from the direct time-integration of Eqs.~\eqref{QRslow}-\eqref{Tslow} starting from asymmetric initial condition. We find two different phases, one with $\kappa=1$ indicating the symmetric solution is realized, and a transition from   $\kappa=1$ to smaller values in the second phase, indicating the symmetric solution is broken. We remark the present of multistability at the boundary between the phases as demonstrated in  Appendix \ref{additional_figures}. Finally, we check the robustness of our results for different $K=8, 16$ in Fig.~\ref{fig7}. Here we find that the value of $\eta$ beyond which the transition occurs is lower with increasing $K$. The transition also sharpens with large $K$. This can be intuitively understood as follows: adding more effectors to
the system creates new pathways to either bypass   a local minimum or escape from it by  turning it into a saddle-point, similar to \cite{FUKUMIZU}.

In the symmetry broken phase, we observe an emergent characteristic of every effector $k$, namely,  it becomes strongly correlated with a specific target gene, say $i$,
while   acquiring a negative correlation with (or decorrelating from) the remaining genes $j\neq i$, as  seen in Fig.~\ref{fig2} \textbf{(f)}. This symmetry-breaking process, however, could also occur
via a sequence of escapes from the symmetric subspace in an order following the
reductions of the covariances $R_{kj}$ \cite{DavidPRE1995}. Similar behavior is observed for $\boldsymbol{T}$ and $\boldsymbol{Q}$, namely, $T_{ij}$ ($Q_{ij}$) decays to negative values  but $T_{ii}$ ($Q_{ii}$) becomes positive, resulting in a significant gap between their diagonal and off-diagonal elements, as shown in Fig. \ref{fig9} of Appendix \ref{additional_figures}. Since  the  target-gene interaction matrix $\boldsymbol{W}$  has no negative elements, such anti-correlations are  emergent properties of self-regulation. 

 \section{Discussion}
\label{sec:discussion}
In this paper, we explain robustness  as the  emergence of collective  correlations  between the intracellular components. In our framework, these non-trivial correlations result from  
a broken (permutation) symmetry   of the effectors, where each  effector becomes positively correlated to only one target gene, while being decorrelated to the remaining ones. The symmetry-broken phase giving rise to robustness  is not a lower-variance state in the naive noise-suppression sense. It is a state in which fluctuations become structured correlations  through mutual coordination between   effectors and target genes. At a mechanistic level, we propose that it is necessary for a receiving cell to adjust
its own sensitivity to non-instructive fluctuations  by adaptively modifying the interpretation of the latter until a stable correspondence between effectors and target genes is established. At a macroscopic level, the consolidation of  cellular  adaption to signal variations is achieved via a ``cooperative'' mechanism  in a similar way to \cite{Inoue}.

Furthermore, we remark that  
it remains challenging to reliably disentangle the effect of correlations from  interactions in complex genetic networks, especially when temporal resolution is limited and one must rely on pseudotime inference or pooled single-cell measurements. In fact, the correlation between a pair of transcriptional factors is determined not only by their direct interactions but also by the dynamic state of the entire network to which they belong. This is why, for comparison with experiments, it is important to have  predictions from such a model, like ours, that exclude the antagonistic correlations generated by assuming mutually repressive interactions. 

The approach based on optimal control~\cite{Pezzotta}
does not account for the slow adaptation of  effectors and treats them as optimal functions of  target genes, thus allowing  the latter  to reach a  predertemined state $\boldsymbol{s}_*$. 
In our case the effectors need to adapt to a 
\emph{moving} ``reference'' $\boldsymbol{s}_*$ among the multitude of $\boldsymbol{s}_*$  
 which the target genes can converge to. Once a particular $\boldsymbol{s}_*$ is attained through adaptation, it remains stable. In control-theory language, our approach is closely related to an adaptive closed-loop controller rather than  an integral feedback control considered in~\cite{Russo} for homeostasis.  
  The distinction is that our
robustness is not defined as  returning to   a baseline state but as the emergence of a stable covariance structure linking    signal transducers and target genes. 

Our framework can be easily extended to  incorporate  
various adaptation mechanisms that have not been considered in the present paper, such as  
adaptation  via epigenetic modifications represented by  a threshold $\theta(t)$ for gene expression~\cite{plugers2025,West} or  adaptation based on network motifs~\cite{Ma2009,Qiao} or time-evolving GRNs \cite{Pham_2024, pham2026entropyproductionratestochastically}, or
 a time-dependent adaptation rate  $\eta=\eta(\tau)$.  In the latter case, it would be interesting to  
study a global optimal control problem of  identifying  which $\eta$ value or schedule  provides the optimal reduction of the  mismatch $\epsilon$  within a given time window~\cite{DavidPRL1997, Rattray1999}.  

While our microscopic model does not  describe a specific GRN in detail, it is able to characterize  the behavior of GRNs
with similar  dynamics and statistics of coupling strengths. Moreover, our macroscopic theory is scalable to large signaling systems.
In order to validate our model, it is necessary to use  topological information inferred from data \cite{Pratapa} 
and to calibrate its parameters with experimentally observed  covariations \cite{Gupta}. In particular,  the parameters $\eta$ and $\gamma$ should  be calibrated  so  that a distribution of cell types that is consistent with single-cell
data can be obtained. 

We also leveraged classical online-learning theory to the far-reaching domain of system biology, paving the way to address some fundamental questions, such as the efficiency of adaptation, that is, how well the cell utilizes its energy budget and other resources to maintain a robust differentiation or   trade-offs 
between different regulatory strategies~ \cite{Ramsden2023}.


 
\begin{acknowledgments}
  We thank  Adri\'an Aguirre-Tamaral, Claudio Hern\'andez-L\'opez, Davey Plugers and Riccardo Rao for insightful discussion. TP is supported by the Dutch Institute for Emergent Phenomena  at the
University of Amsterdam under the Research Priority Area \emph{Emergent Phenomena in Society: Polarisation, Segregation and Inequality}. Financial support from the UK Multidisciplinary Centre for Neuromorphic Computing (UKRI982) is gratefully acknowledged (DS).
\end{acknowledgments}


\appendix

\section{Gaussian integrals and the mismatch}
\label{sec:formulation}
First, let us remark that for jointly Gaussian zero-mean variables \(x,y\), we have the following standard identity
\begin{equation}
\big\langle \phi(x)\phi(y)\big\rangle
=
\frac{2}{\pi}
\arcsin\!\left(
\frac{\operatorname{Cov}(x,y)}
{\sqrt{(1+\operatorname{Var}x)(1+\operatorname{Var}y)}}
\right).
\end{equation}

Now we introduce the Gaussian integrals $I_3$ and $I_4$, which, for $u,v,y,z$ being one of the components of either $\boldsymbol{h}$ and $\boldsymbol{s}$, are defined as 
\begin{equation}
    \begin{aligned}
      I_3(u,v,y)&\equiv \Big\langle \phi'(u)v\phi(y)\Big\rangle\\ I_4(u,v,y,z) &\equiv\Big\langle \phi'(u) \phi'(v) \phi(y) \phi(z)  \Big\rangle
    \end{aligned}
\end{equation}
These Gaussian integrals can be computed explicitly as function of $(\boldsymbol{T}, \boldsymbol{Q},\boldsymbol{R})$ as detailed in \cite{DavidPRE1995}. For self-contained presentation, we rewriting them below
\begin{equation}
    I_3(u,v,y) = \frac{2}{\pi}\, \frac{1}{\sqrt{\Lambda_3}} \frac{C_{23}(1+C_{11}) - C_{12}C_{13}}{1+C_{11}}
\end{equation}
where we denote  $u,v,y, z$ by indices $1,2,3, 4$ and 
\begin{equation}
    \Lambda_3 = (1+C_{11})(1+C_{33}) - C^2_{13}
\end{equation}
For example,
\begin{equation*}
\boldsymbol{C}(h_k,h_\ell,s_i)=  \begin{pmatrix}  \displaystyle Q_{kk} & Q_{k\ell} & R_{k i} \vspace*{0.1cm} \\ Q_{\ell k} & Q_{\ell\ell} & R_{\ell i} \\ R_{k i}  &R_{\ell i} & T_{ii} \end{pmatrix}      
\end{equation*}
and for $I_4$
\begin{equation}
    I_4(u,v,y,z) = \frac{4}{\pi^2}\, \frac{1}{\sqrt{\Lambda_4}}~{\rm arcsin}\left(\frac{\Lambda_0}{\sqrt{\Lambda_1 \Lambda_2}}\right)
\end{equation}
with
\begin{equation}
    \begin{aligned}
    \Lambda_4=  &~  (1+C_{11})(1+C_{22}) -C^2_{12} \\
    \Lambda_0 = &~ \Lambda_4C_{34} -C_{23}C_{24} (1+C_{11}) - C_{13}C_{14}(1+C_{22}) \\&~+C_{12}C_{13}C_{24} +C_{12}C_{14}C_{23}
    \\
    \Lambda_1 = &~ \Lambda_4C_{33} -C^2_{23} (1+C_{11}) - C^2_{13}(1+C_{22}) \\&~+2C_{12}C_{13}C_{23} 
     \\
    \Lambda_2 = &~ \Lambda_4(1+C_{44}) - C^2_{24}(1+C_{11}) -C^2_{14}(1+C_{22})\\&~+2C_{12}C_{14}C_{24}
    \end{aligned}
\end{equation}
We here provide the explicit forms of the functions $I_{k\ell}$ and $\tilde{I}_{k\ell}$ in Eq. \eqref{QRslow}, both are sum of
as follows:
\begin{equation}
\begin{aligned}
I_{k\ell} & = \sum_{i =1}^K \Big[I_3(h_k,h_\ell,s_i) + I_3(h_\ell,h_k,s_i)\Big]\\ &- \sum_{j =1}^K \Big[I_3(h_k,h_\ell,h_{j}) + I_3(h_\ell,h_k,h_{j})\Big]\\
  \tilde{I}_{k \ell} &= \sum_{i,j =1}^K I_{4}(h_k,h_\ell,s_i,s_j)  +\sum_{i,j =1}^K I_{4}(h_k,h_\ell,h_i,h_j)\\&- 2 \sum_{i,j =1}^K I_{4}(h_k,h_\ell,h_{i},s_j)
\end{aligned}
\end{equation}
The mismatch can be given in terms of the macroscopic variables  as follows:
\begin{equation}
\begin{aligned}\epsilon&= \pi^{-1}\sum_{i,k} \left\{ {\rm arcsin }\frac{Q_{ik}}{\displaystyle \sqrt{1+Q_{ii}}\sqrt{1+Q_{kk}}}\right.\\
    &+{\rm arcsin }\frac{T_{ik}}{\displaystyle \sqrt{1+T_{ii}}\sqrt{1+T_{kk}}}\\&-2 \left. {\rm arcsin }\frac{R_{ik}}{\displaystyle \sqrt{1+Q_{ii}}\sqrt{1+T_{kk}}}\right\}\end{aligned}
\end{equation}


\section{Derivation of Eqs.~\eqref{symmetric_with_noise}-\eqref{symmetric_no_noise}}
\label{sec:derivation_sym}

On the symmetric manifold with $Q_{ii}=Q_{ij}=q$; $R_{ii}=R_{ij}= r$;  $T_{ii}= u$ and $T_{ij}=v$, for $\eta\ll 1$,  the slow flow in Eq.~\eqref{QRslow} reduces to
\begin{equation}
\begin{aligned}
\dot q &= \eta\big[\,2K(\phi-\theta)\,\big]\\
\dot r &= \eta\big[\,\psi-K\beta\,\big],
\end{aligned}
\label{qr_symmetric_flow}
\end{equation}
where
\begin{equation}
\begin{aligned}
\phi&:=\frac{2}{\pi}\frac{r}{(1+q)D}\,,\quad 
\theta:=\frac{2}{\pi}\frac{q}{(1+q)L}\,,\quad  
\beta:=\frac{2}{\pi}\frac{r}{(1+q)L},\\
\psi&:=\frac{2}{\pi}\,
\frac{(1+q)\big(u+(K-1)v\big)-Kr^2}{(1+q)D}\\D&=\sqrt{(1+q)(1+u)-r^2},
\qquad
L=\sqrt{1+2q}.
\end{aligned}
\label{D_and_L}
\end{equation}
On the other hand, symmetric solution to the  Eq.~\eqref{Tslow} reads 
\begin{equation}
\begin{aligned}
  v &=  \frac{\sigma^2}{2}~\frac{cw}{\mu(1+cw)}
+\sqrt{\frac{2}{\pi}}~\frac{\gamma}{\mu}~\frac{r}{\sqrt{1+q}}\\ u&=v + \frac{\sigma^2}{2}~\frac{1}{1+cw} \\
\mu &:=1-cw(K-1)
\end{aligned}
\label{uv_symmetric_flow}
\end{equation}
The fixed-point conditions, $\phi=\theta$ and $
\psi=K\beta$, once combined yileds
\begin{equation}
    r^2=\frac{q^2}{1+q}(1+u) \,,\quad u+(K-1)v=\frac{Kr^2}{q}
    \label{fixed_point_symmetric_flow}
\end{equation}
Substituting Eq.~\eqref{uv_symmetric_flow} into the above equation, we arrive at
\begin{equation}
r^2=q^2\kappa^2\,,\quad \kappa^2:=
1+\frac{\sigma^2}{2\mu}
\left[
\frac{1-cw(K-2)}{1+cw}-\frac{1}{K}
\right]
\end{equation}
Choosing the positive branch relevant for positive overlaps $r$ we obtain Eq.~\eqref{symmetric_with_noise}.

Let $y:=\sqrt{1+q}$, equation~\eqref{q_symmetric_with_noise} then becomes
\begin{equation}
\kappa^2 y^3
-
\frac{\gamma\alpha\kappa}{\mu}y^2
-
\left(\kappa^2+\frac{\sigma^2}{2K\mu}\right)y
+
\frac{\gamma\alpha\kappa}{\mu}
=0.
\end{equation}
For $\sigma\rightarrow 0$, $\kappa\rightarrow 1$,  this equation reduces to
\begin{equation}
\kappa(y^2-1)\left(\kappa y-\frac{\gamma \alpha }{\mu}\right)=0   
\end{equation}
To have a non-trivial symmetric solution $q=y^2-1$,  we need such a real root  $y\geq 1$  of this cubic equation  \begin{equation}
    y= \frac{2}{\pi}~\frac{\gamma^2}{\kappa\big[1-cw(K-1)\big]^2}\,,
\end{equation}
from which we obtain Eq.~
\eqref{symmetric_no_noise}.

\begin{figure*}
    \centering
\includegraphics[width=0.96\linewidth]{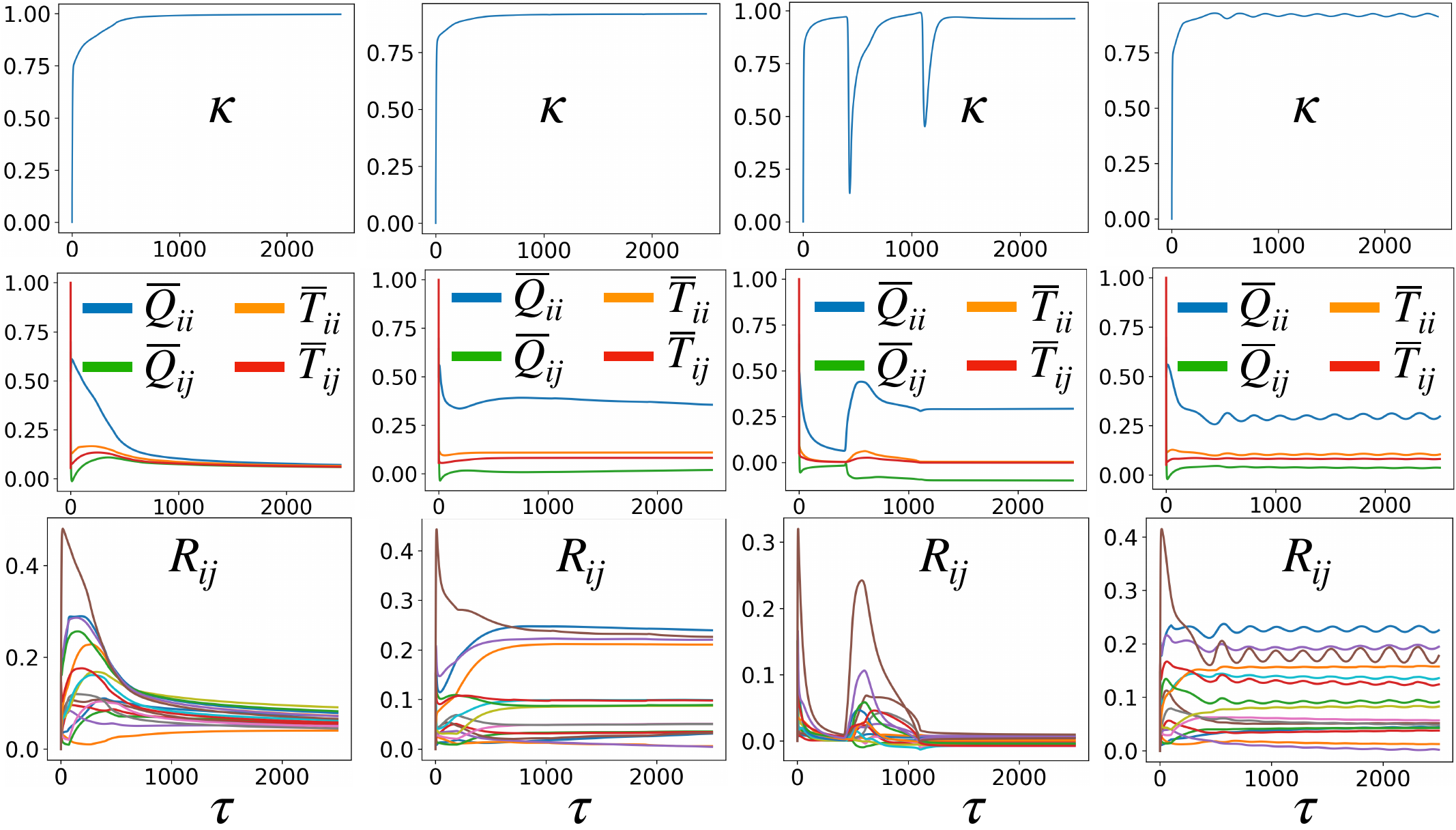}
\caption{\textbf{Multistability}. From top to bottom: the ratio     $\kappa = \bar{r}/\bar{q}$ with $\bar{r}\equiv K^{-2}\Big[\sum_{i} R_{ii} +\sum_{i\neq j} R_{ij}\Big]$ and $\bar{q}\equiv K^{-2} \Big[\sum_{i} Q_{ii} +\sum_{i\neq j} Q_{ij}$\Big], all covariances  $R_{ii}$ and $R_{ij}$, the averaged diagonal  and off-diagonal  overlaps $(\overline{Q}_{ii}, \overline{T}_{ii})$ and $(\overline{Q}_{ij}, \overline{T}_{ij})$, respectively.
Here  $\eta=\gamma =0.9$,   $c = 0.2$, $\sigma = 0.01$, $K=4$, and $\boldsymbol{W}$ is an \emph{asymmetric} matrix of uncorrelated elements drawn from $\mathcal{U}(0,1)$ -- the uniform distribution in $(0,1)$. We use symmetric initial conditions:  $Q_{ii}=Q_{ij}= 0.5$, $R_{ii}=R_{ij} =10^{-6}$, $T_{ii} = 1+\sigma^2/(2*(1+c \langle W\rangle))$, $T_{ij} = 1$.}
\label{fig8}
\end{figure*}

\section{Onset of specialisation}
\label{sec:Jacobian}
To study the incipient of specialization, in which a dominant overlap $R$ between a given 
target gene and its corresponding effector    can be distinguished from a secondary overlap $S\neq R$ with all other  effectors, we consider the permutation-equivariant ansatz
\begin{equation}
\begin{aligned}
Q_{ii}&=Q\,,\quad Q_{ij} = C,
\\
R_{ii}&=R\,,\quad R_{ij} =S,
\\
T_{ij}&=U\,,\quad T_{ij} =V.
\end{aligned}
\end{equation}
Next we expand this ansatz round the symmetric solution given in Eqs.~\eqref{symmetric_with_noise}-\eqref{q_symmetric_with_noise} with $\varepsilon\ll 1$
$$Q=q+\varepsilon\delta_q\,,
C=q+\varepsilon\delta_c\,, R=r+
\varepsilon\delta_r\,,
S=r+\varepsilon\delta_s$$
and check  the linearised slow dynamics Eq.~\eqref{QRslow}:
\begin{equation}
\frac{d}{d\tau}
\begin{pmatrix}
\delta_q-
\delta_c\\
\delta_r-
\delta_s
\end{pmatrix}
=\eta
\mathbb{J}_-
\begin{pmatrix}
\delta_q-
\delta_c\\
\delta_r-
\delta_s 
\end{pmatrix} + O(\eta^2)
\end{equation}
where $\mathbb{J}_-$ is the transverse sector of the full Jacobian under the permutation-equivariant ansatz. In \cite{DavidPRE1995}, at the leading order, $\delta_q=\delta_c$ due to geometric reason, while $\delta_r\neq \delta_s$. In our case, $\delta_r\neq \delta_s$ and $\delta_q\neq \delta_c$ happens at the same time. The expression of $\mathbb{J}_-$, with $D$ and $L$ given in Eq.~\eqref{D_and_L}, then reads 
\begin{equation}
    \mathbb{J}_-=\eta
\begin{pmatrix}
a & e\\
g & m
\end{pmatrix}
\end{equation}
where  
\begin{equation}
\begin{aligned}
a&= \frac{4}{\pi}
\left[ \frac{q}{(1+q)L^3}+ (K-1)
\frac{1+3q+q^2}{(1+q)^2L} - \frac{1}{(1+q)^2L}\right]
\\
&\quad -\frac{4}{\pi}~Kr
\left[
\frac{1}{(1+q)^2D}
+
\frac{1+u+(1+q)u_q}{2(1+q)D^3}
\right]
\end{aligned}
\end{equation}
\begin{equation}
    e
=
\frac{4}{\pi}
\left[
\frac{2-K}{D(1+q)}
-\frac{r\big(K(1+q)u_-+2(K-2)r\big)}{2(1+q)D^3}
\right]\,,
\end{equation}
\begin{equation}
    \begin{aligned}
m
&=
\frac{2}{\pi}
\left[
\frac{(1+q)u_- -2r}{D(1+q)}
-
\frac{A[(1+q)u_- -2r]}{2D^3(1+q)}
\right]
\\
&\quad+
(K-1)\frac{2}{\pi}
\left[
\frac{v_-}{D}
-
\frac{B[(1+q)u_-+2r]}{2D^3(1+q)}
\right]\\&\quad
-\frac{2}{\pi}~\frac{1}{L(1+q)}
+
\frac{2}{\pi}~\frac{L(K-1)}{(1+q)}.
\end{aligned}
\end{equation}

\begin{equation} 
\begin{aligned}
g&=(K-1)\frac{2}{\pi}
\left[
\frac{v+\Delta}{(1+q)D}
-
\frac{B}{(1+q)^2D}
-\frac{B(1+u+\Delta)}{2(1+q)D^3}\right]
\\
&\quad
+ \frac{2}{\pi}
\left[
\frac{u+\Delta}{D(1+q)}
-
\frac{A}{(1+q)^2D}
-
\frac{A(1+u+\Delta)}{2(1+q)D^3}
\right]
\\
&\quad
+
r\frac{2}{\pi}
\left[
\frac{1}{(1+q)^2L}
+
\frac{1}{(1+q)L^3}
\right]
-
\frac{2}{\pi}
\frac{qr(K-1)}{(1+q)^2L}.
\end{aligned}
\end{equation}
and we have introduced
\begin{equation}
\begin{aligned}
A &:=u(1+q)-r^2\,,\quad 
B:=v(1+q)-r^2\\
    \Delta &= -rC(1+cw)\,,\quad  C=
\frac{\sqrt{\frac{2}{\pi(1+q)}}~\frac{\gamma}{\mu}}{1+cw}\\  u_-:&= C[1+cw-2cw(K-1)]\,,\quad v_-:
=C(cw-1)\,.
\end{aligned}
\end{equation}

\section{Additional Results}
\label{additional_figures}
Here  we first show in Fig. \ref{fig8} the multistability that occurs at the boundary between the symmetric and symmetry-broken phases. Here we find that not only the symmetric and asymmetric solutions coexist, but also a multitude of distict  (a)symmetric solutions.

Next,  to  characterize the change in the effectors' behavior due to the symmetry-breaking mechanism, 
we introduce the following measure:
\begin{equation}
\Delta Q = \overline{Q}_{ii} -\overline{Q}_{ij}\,,\, \overline{Q}_{ii}:=\frac{1}{K}\sum_{i=1}^K Q_{ii} \,,\,  \overline{Q}_{ij}:=\frac{2 \sum_{i<j}Q_{ij}}{K(K-1)}
\label{gap}
\end{equation}
So $\Delta Q$ is the \emph{averaged} gap between diagonal and non-diagonal elements of  $\boldsymbol{Q}$. 
In Fig.~\ref{fig8} we observe  $\Delta Q$ gradually increases from zero to significantly higher values  as the adaptation rate  $\eta$ crosses  a critical point.

\begin{figure}[t]
    \centering
\includegraphics[width=0.76\linewidth]{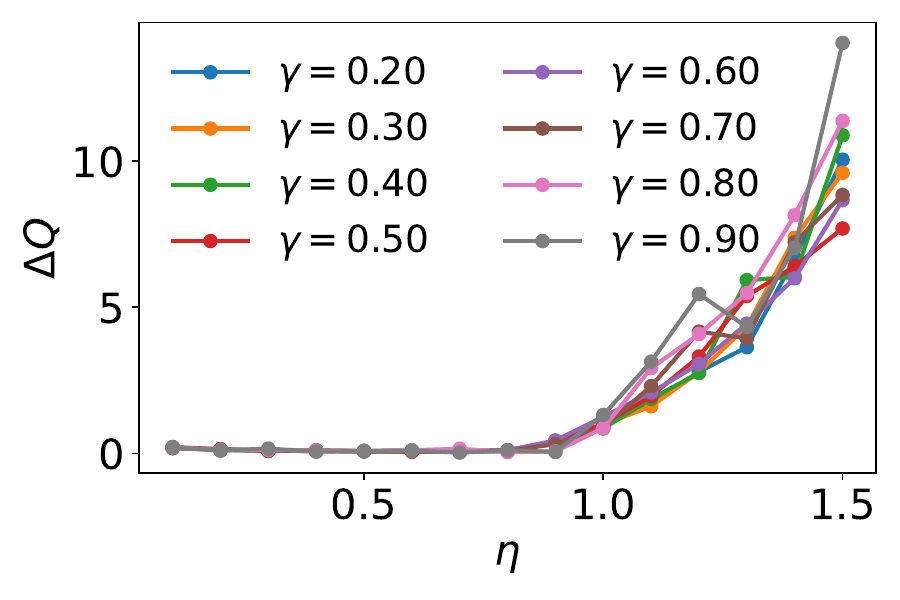}
\caption{Dependence of $\Delta Q$ on $\eta$ for various $\gamma$. Here  $c = 0.2$, $\sigma = 0.01$, $K=4$, and $\boldsymbol{W}$ is an \emph{asymmetric} matrix whose uncorrelated elements  are drawn from $\mathcal{U}(0,1)$ -- the uniform distribution in the interval $(0,1)$. Initial condition for $\boldsymbol{Q}$ and $\boldsymbol{T}$  is specified by $Q_{ii}$ and $T_{ii}$, both  drawn from   $\mathcal{U}(0,1)$,  while $Q_{ij}$ and $T_{ij}$ -- from  the uniform distribution in the interval $10^{-6}\times(0,1)$. All elements of  $\boldsymbol{R}(\tau =0)$ are drawn from the uniform distribution  $10^{-6}\times\mathcal{U}(0,1)$.}
\label{fig9}
\end{figure}


\newpage 

\bibliography{sample}
 \end{document}